\documentclass{optica-article}

\journal{opticajournal} 

\articletype{Research Article}

\usepackage{lineno}

\begin{document}

\title{Sub-shot-noise mid-infrared photothermal measurements using twin beams}

\author{Kazuki Hashimoto,\authormark{1,2,3,*} Takuro Ideguchi,\authormark{1,2,3}}

\address{\authormark{1}Institute for Photon Science and Technology, The University of Tokyo, Tokyo, 113-0033,Japan\\
\authormark{2}Department of Advanced Materials Science, Graduate School of Frontier Sciences, The University of Tokyo, Tokyo, 277-8561,Japan\\
\authormark{3}RIKEN Center for Advanced Photonics, RIKEN, Saitama, 351-0198, Japan}

\email{\authormark{*}k-hashimoto@ipst.s.u-tokyo.ac.jp} 


\begin{abstract*} 
Mid-infrared photothermal (MIP) measurement probes refractive-index changes induced by mid-infrared (MIR) absorption using visible or near-infrared light and has recently emerged as a powerful modality for MIR spectro-imaging in the life sciences. Reducing measurement noise allows the probe power to be lowered without compromising the signal-to-noise ratio (SNR), thereby mitigating the risk of sample damage. However, even in the absence of classical noise, the SNR of conventional MIP measurements using a classical coherent probe is fundamentally limited by shot noise. Here, we demonstrate sub-shot-noise MIP measurement using twin beams generated through seeded optical parametric amplification. Balanced detection of the twin beams suppresses the noise by more than 5 dB below the shot-noise level established using a coherent beam. As a proof-of-principle demonstration, we measure the MIP signal from liquid water using balanced twin-beam detection and achieve an SNR enhancement of more than 2 dB over the balanced coherent-beam detection. 
\end{abstract*}

\section{Introduction}
Photothermal sensing is an optical measurement technique in which a probe beam detects refractive-index changes induced by the photothermal effect arising from pump-beam absorption, and has found widespread applications in trace analysis of various targets, including colorless organic compounds~\cite{Long1976}, microparticles~\cite{Wu1991}, single nanoparticles~\cite{Boyer2002}, and single molecules~\cite{Gaiduk2010}. Recently, photothermal sensing has been applied for mid-infrared (MIR) microspectroscopy~\cite{Yin2025_2,Ideguchi2025}, particularly for life science applications. By leveraging the diffraction-limited spatial resolution of visible or near-infrared wavelengths together with high-performance optical detection in this spectral range, MIR photothermal (MIP) techniques enable high-spatial-resolution microscopy sensitive to fundamental vibrational modes. 

Reducing measurement noise in MIP allows lowering probe power without compromising the signal-to-noise ratio (SNR), thereby mitigating the risk of light-induced sample damage. The sensitivity of MIP measurements is affected by various sources of technical noise, including detector noise and the relative intensity noise of the probe laser. Even when these technical noises are sufficiently suppressed, the measurement remains fundamentally subject to photon shot noise, which is determined by the square root of the detected mean photon number. Shot-noise-limited or near-shot-noise-limited performance has been demonstrated in various MIP configurations, including wide-field~\cite{Ishigane2023,Bai2019,Paiva2022}, confocal interferometric~\cite{Zhang2021}, and balanced-detection~\cite{Waclawek2019} implementations. At the shot-noise limit, further SNR improvement with a classical coherent probe requires higher probe power, inevitably increasing the risk of sample photodamage.

The shot-noise level can be further reduced by employing quantum light sources~\cite{Chekhova2015,Andersen2016,Meda2017,Lawrie2019}. Among quantum light sources, twin beams~\cite{Heidmann1987} exhibiting intensity-difference squeezing have been shown to improve measurement SNR in a wide range of applications, including imaging~\cite{Boyer2008,Brida2010,Samantaray2017,SabinesChesterking2019}, absorption measurements~\cite{Jakeman1986,Moreau2017,Mueller2020,Li2021}, and thermoreflectance sensing~\cite{Ather2023}, to name a few. Their advantages have also been extended to various spectroscopic modalities, such as absorption spectroscopy~\cite{SoutoRibeiro1997,Whittaker2017,Matsuzaki2022}, dual-comb spectroscopy~\cite{Wan2025}, time-resolved THz spectroscopy~\cite{Adamou2025}, stimulated Raman scattering spectroscopy~\cite{Xu2022}, and stimulated Brillouin scattering spectroscopy~\cite{Li2022}. Recently, Fu et al. have applied one of the twin beams generated via four-wave mixing (FWM) in Rb vapor to a photothermal probe to demonstrate quantum-enhanced visible photothermal microscopy~\cite{Fu2026}. However, to the best of our knowledge, quantum-enhanced MIP measurements have not yet been demonstrated.

In this study, we demonstrate sub-shot-noise MIP measurements using pulsed twin beams generated via seeded optical parametric amplification (OPA) in a periodically poled lithium niobate (PPLN) waveguide. Balanced detection of the twin beams yields a noise reduction of >5 dB compared to the coherent-beam case. One of the twin beams is used as a probe for the MIP effect induced by a quantum-cascade laser (QCL) with 2 MHz intensity modulation, while the other serves as a reference in the balanced detection scheme. Using this configuration, we perform a proof-of-principle experiment by measuring the MIP signal from liquid water, achieving an SNR enhancement of more than 2 dB over the balanced-detection coherent-probe case.

\section{Results}
\subsection*{Principle of sub-shot-noise MIP}

\begin{figure}[htbp]
\centering
\includegraphics[width=1\linewidth]{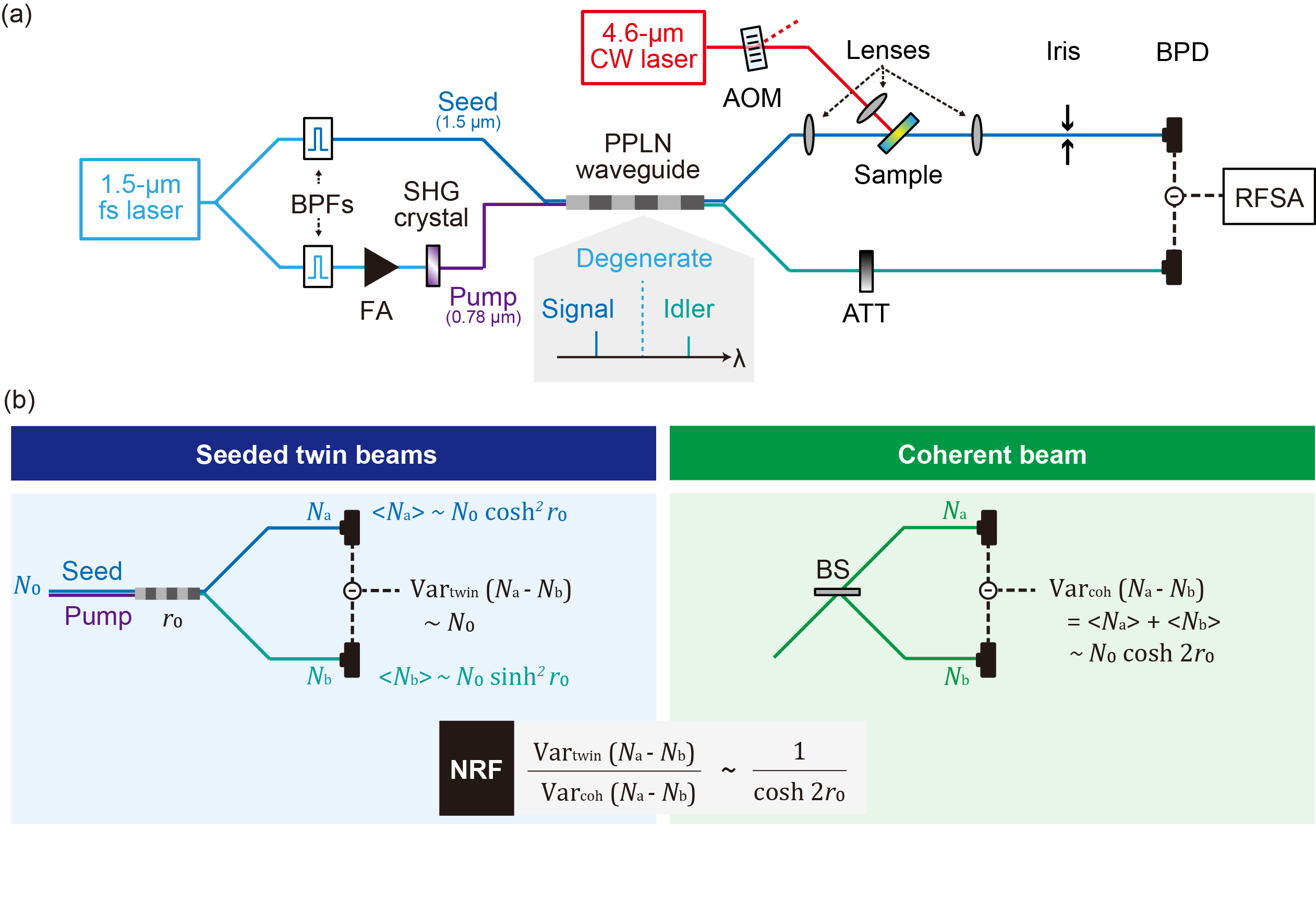}
\caption{Principle of sub-shot noise MIP measurement with twin beams.(a) Schematic of sub-shot-noise MIP measurement system. BPF: band-pass filter, FA: fiber amplifier, SHG: second-harmonic generation, PPLN: periodically poled lithium niobate, CW: continuous wave, AOM: acousto-optic modulator, ATT: attenuator, BPD: balanced photodetector, RFSA: radio-frequency spectrum analyzer. (b) Noise reduction with seeded twin beams. When the overall photon number incident on the BPD is the same, the variance is reduced to around $1/\cosh(2r_0)$ relative to that of a coherent beam, where $r_0$ is the squeezing parameter. $N_0$: mean photon number of a seed beam, $N_{a(b)}$: photon-number operators of signal (idler), $\mathrm{Var}(\cdot)$: variance, $\langle \cdot \rangle$: mean. NRF: noise reduction factor, BS: beamsplitter}
\label{fig:schematic}
\end{figure}

The concept of sub-shot-noise MIP measurement is illustrated in Fig.~\ref{fig:schematic}(a). The detailed experimental implementation is shown in Methods and Supplementary Note 1. A 1.5-µm femtosecond laser (fundamental beam) is split into two paths, providing the seed and pump for the OPA. In the pump path, the fundamental beam is spectrally filtered to a bandwidth of 0.2 nm using a fiber Bragg grating (FBG) and subsequently amplified by a fiber amplifier. The amplified beam is then directed to a PPLN crystal to generate a 0.78-µm beam via second-harmonic generation (SHG), which serves as the OPA pump. In the seed path, the fundamental beam is filtered to a bandwidth of 0.2 nm using a custom-built spectral filter and passes through a delay line to control the temporal delay between the pump and seed beams. The central wavelength of the seed is slightly detuned from the degenerate wavelength of the OPA process. The pump and seed beams are then coupled into a PPLN waveguide, where a phase-insensitive OPA generates seeded twin beams (signal and idler) at the output. The waveguide confines the light to a small mode area over a long interaction length, enabling efficient generation of squeezed light or twin beams~\cite{Eto2007,Harder2016,Hirota2026}. 

Balanced detection of the twin beams enables sub-shot-noise measurements, as illustrated in Fig. ~\ref{fig:schematic}(b). Assuming that the twin beams are generated via a phase-insensitive OPA seeded by a coherent beam with a mean photon number $N_0$, the mean photon numbers of the signal and idler are denoted as $\langle N_a \rangle\approx N_0\cosh^2r_0$ and $\langle N_b \rangle\approx N_0\sinh^2r_0$, respectively, with a total photon number of $N_0\cosh2r_0$, where $r_0$ is the squeezing parameter. $r_0$ characterizes the strength of the parametric amplification and determines the signal gain as $\cosh^2r_0$. In the absence of loss, the variance of the photon number difference $\mathrm{Var}(N_a-N_b)$ is approximately equal to $N_0$. In contrast, for a coherent beam with the same total mean photon number, $\mathrm{Var}(N_a-N_b)$ is equal to $N_0\cosh2r_0$, corresponding to the shot-noise limit. The noise reduction factor~\cite{Heidmann1987} (NRF), defined as the ratio of these variances for twin beams and coherent beam (or equivalently, the ratio of these variance for twin beams and the total mean photon number), is given by $1/\cosh(2r_0)$.  This indicates that balanced detection of the twin beams enables measurements below the shot-noise level. For example, when $r_0$=1.5, the NRF is approximately 0.1 under ideal conditions. A detailed theoretical description of the noise reduction using seeded twin beams is provided in Supplementary Note 2.

The twin-beam detection scheme is applied to the MIP measurement system. A continuous-wave (CW) QCL operating at 4.64 µm (2153 cm$^{-1}$) is used as the MIP pump source, with its intensity modulated by an acousto-optic modulator (AOM). The pump beam is focused onto a liquid sample contained in a sample cell composed of two CaF$_2$ windows separated by a spacer, thereby inducing the MIP effect. The signal beam is focused into the sample to probe the resulting MIP response. After passing through the sample, the signal beam is recollimated and directed through an iris placed at the Fourier plane. The MIP response periodically changes the spatial distribution of the probe beam at this plane. By transmitting only part of that distribution, the iris converts this change into a modulation of the transmitted probe power at the pump modulation frequency. The transmitted signal beam is directed to one port of a high-quantum-efficiency balanced photodetector (BPD). The idler beam, serving as a reference, is directed to the other port after appropriate power adjustment for balancing. The differential output of the BPD is analyzed using a radio-frequency spectrum analyzer (RFSA). Owing to the noise reduction enabled by the twin beams, the noise floor is reduced below the shot-noise level, resulting in an enhanced SNR.

\subsection*{Twin-beam generation}
\begin{figure}[htbp]
\centering
\includegraphics[width=1\linewidth]{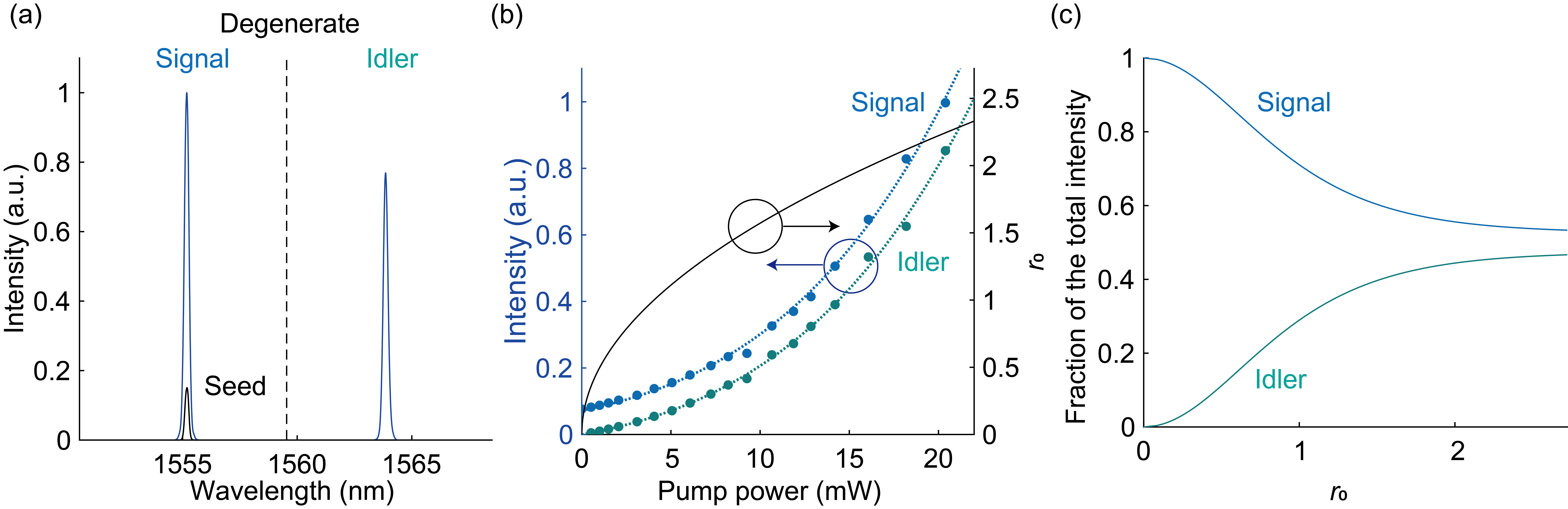}
\caption{Seeded twin-beam evaluation. (a) Twin-beam spectra (blue) generated via the seeded OPA with a pump power of 14.2 mW. The black line shows the seed beam spectrum measured without the pump beam. (b) The signal (blue) and idler (blue-green) spectral intensity versus the pump power. The dotted blue and blue-green indicate the fitting curves for signal and idler, respectively. The black line shows $r_0$ versus pump power obtained from the idler fitting results. (c) Signal (blue) and idler (blue-green) fractions of the total intensity as a function of $r_0$.}
\label{fig:twin_beam}
\end{figure}

First, we characterize the twin beams generated by the seeded OPA. Figure~\ref{fig:twin_beam}(a) shows twin-beam spectra generated via seeded OPA with a pump power of 14.2 mW and the seed spectrum, measured using an optical spectrum analyzer (OSA) with a spectral resolution of 0.05 nm. The OPA pump is generated via SHG of the fundamental beam centered at 1559.5 nm with a bandwidth of 0.2 nm, while the OPA seed had a center wavelength of 1555.2 nm with a bandwidth of 0.2 nm. Detailed characterizations of the fundamental and SHG spectra, as well as the SHG conversion efficiency, are provided in Supplementary Note 3. The central wavelengths of the signal and idler spectra are 1555.2 and 1563.9 nm, respectively, which are slightly detuned from the degenerate wavelength. The frequency detunings of the twin beams are chosen to be sufficiently large to enable clear spectral separation of the signal and idler by an optical filter, while ensuring that the signal wavelength remains within its passband. The twin-beam spectra are contained fully within the spontaneous parametric down-conversion (SPDC) bandwidth (see Supplementary Note 4).

Figure~\ref{fig:twin_beam}(b) shows the spectral intensities of the signal and idler beams as a function of the OPA pump power, together with the fitting results. The fitting function for the idler is $y = A_i\mathrm{sinh}^2 (B \sqrt{x})$, where $A_i$=0.0389,  and $B$=0.497 are the fitting coefficients, assuming that the mean photon number of the seed beam is sufficiently larger than 1. The signal results are fitted by $y = A_s\mathrm{sinh}^2 (B \sqrt{x})+C_s$ using the obtained coefficient $B$, where $A_s$=0.0432  and $C_s$=0.0722 are the fitting coefficients. As the pump power increases, both the signal and idler intensities increase exponentially. The squeezing parameter $r_0$ can be estimated from $B \sqrt{x}=r_0$, as shown in Fig.~\ref{fig:twin_beam}(b). Using these results, the relative signal and idler fractions of the total intensity as a function of $r_0$ are shown in Fig.~\ref{fig:twin_beam}(c). As $r_0$ increases, the signal-to-idler ratio asymptotically approaches 1:1. These curves are used to estimate the mean photon numbers per pulse of the signal and idler beams from the total average power of the twin beams. The detuning of the twin beams can be adjusted by changing the center frequency of the seed beam using a custom-built spectral filter. The twin-beam spectra seeded at 1553.9 nm, together with the corresponding characterization results, are presented in Supplementary Note 4.

\subsection*{Twin-beam noise reduction}
We next evaluate the intensity-difference noise reduction of the seeded twin beams by directing the two beams separated by the BPF directly into the respective input ports of the BPD. In the present experimental setup, the light source operates at a repetition rate of 216.7 MHz, whereas the photodetector has a bandwidth of 10 MHz. Consequently, the detector does not resolve individual pulses but instead measures low-frequency photocurrent fluctuations below the repetition frequency. In view of the classical-noise peak observed near 1 MHz, and to avoid excess classical noise at lower frequencies, both the MIP pump modulation frequency and the frequency band used to evaluate the probe shot noise are selected in the several-MHz range within the 1–10 MHz detection bandwidth. To characterize the noise reduction in the twin beams, the noise power is measured at 5 MHz using an RFSA under the zero-span mode with a resolution bandwidth of 1 MHz and a video bandwidth of 1 kHz. 

\begin{figure}[htbp]
\centering
\includegraphics[width=1\linewidth]{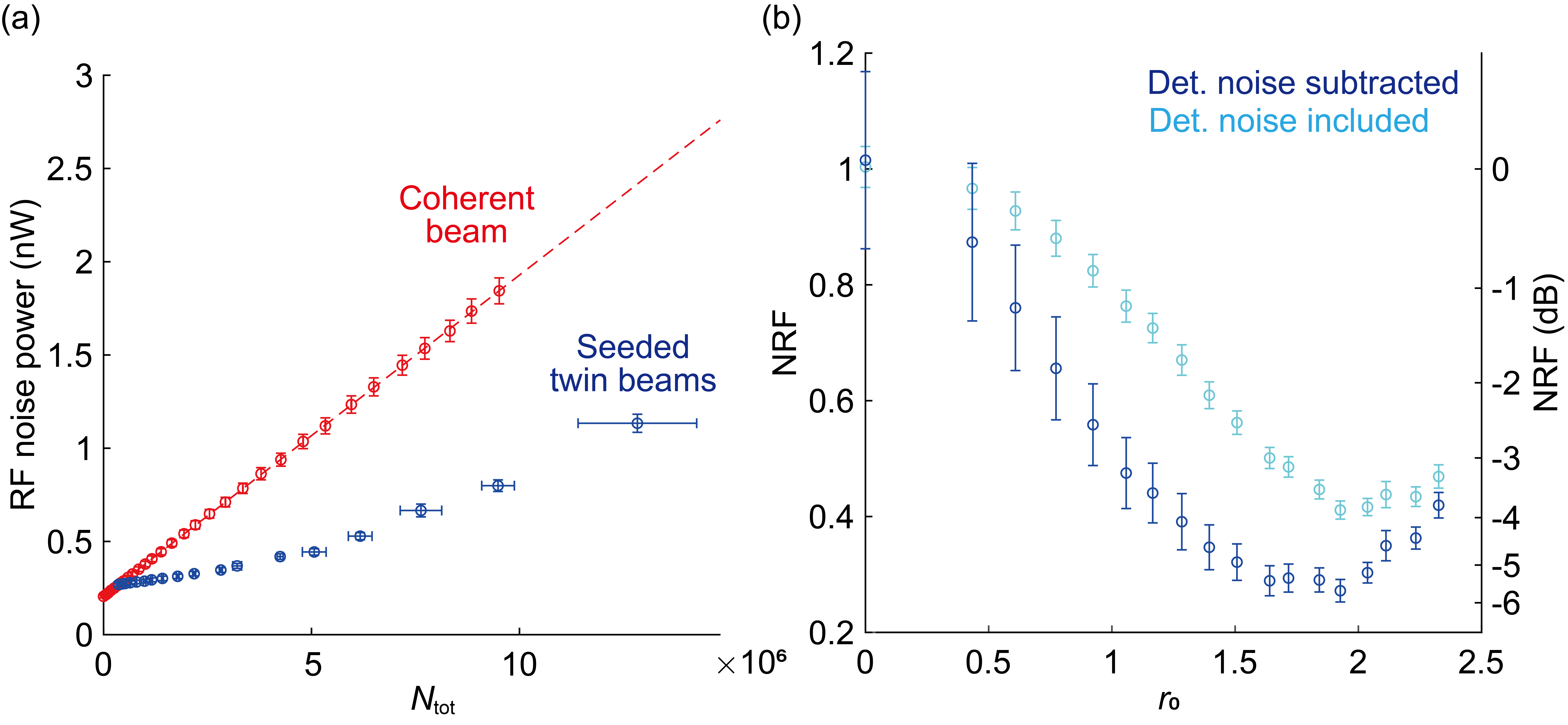}
\caption{Noise reduction achieved by seeded twin beams. (a) RF noise power of the twin beams with $N_{\mathrm{0}}=$ 3.6 x 10$^5$ measured with balanced detection (blue), compared with the shot-noise level of a coherent beam (red). The red dashed line represents a linear fit to the shot-noise level of the coherent beam. The x- and y-axis error bars represent the standard deviations of $N_{\mathrm{tot}}$ and the standard deviation of the RF noise power over time, respectively. (b) Corresponding NRF as a function of $r_0$ with (blue) and without (sky-blue) the detector noise subtraction. The y-axis error bars represent the standard deviations of the NRFs obtained by propagating the standard deviations of the RF noise powers shown in (a).}
\label{fig:NRF_twin}
\end{figure}

Figure~\ref{fig:NRF_twin}(a) shows the RF noise power as a function of the total mean photon number per pulse $N_{\mathrm{tot}}$ for both the coherent beam and the twin beams. The result for the coherent beam corresponds to the detection of the seed beam at a wavelength of 1555.2 nm, where the seed beam passes through the BPF and is detected by one port of the BPD (single-port detection). Strictly speaking, a small fraction of the light is reflected by the BPF and is also detected by the other port of the BPD. $N_{\mathrm{tot}}$ is calculated from the average optical power measured before the BPF using a power meter. As indicated by the fitting results in Fig.~\ref{fig:NRF_twin}(a), the RF noise power of the coherent beam exhibits a linear dependence on $N_{\mathrm{tot}}$, confirming that the measurement is limited by shot noise, while the y-intercept corresponds to the detector noise floor. Since the RF noise power is determined by $N_{\mathrm{tot}}$, replacing the BPF with a beamsplitter (BS) and switching from single-port detection to balanced detection yields the same result in the ideal case. We confirm this, as discussed in detail in Supplementary Note 5. The RF noise power of the intensity difference between the twin beams is then measured under the same conditions. The seed photon number per pulse $N_{\mathrm{0}}$ for the OPA is 3.6 × 10$^5$. The signal and idler beams are separated by the BPF and detected at the respective ports of the BPD. $N_{\mathrm{tot}}$ for the twin beams is increased by increasing $r_0$. The twin-beam plots are below the shot-noise level of the coherent beam, demonstrating sub-shot-noise characteristics. 

The corresponding NRFs are calculated by dividing the RF noise power of the twin-beam intensity difference by that of the shot-noise reference, using the measurements described above. Figure~\ref{fig:NRF_twin}(b) shows the NRFs as a function of $r_0$. Assuming that the photocurrent noise originating from the probe and the electronic noise of the detector are statistically independent, their noise powers add linearly. Therefore, in addition to the raw NRF, which includes detector noise, we also plot the detector-noise-corrected NRF. The corrected NRF is calculated by subtracting the electronic-noise power from both the measured twin-beam difference-noise power and the shot-noise reference. The NRFs first decrease with increasing gain, then turn upward. The best NRF with detector-noise subtraction is -5.7 dB at $r_0=$1.9. Similar evaluations of the sub-shot-noise characteristics,  performed for $N_{\mathrm{0}}=$ 7.3 × 10$^5$, 1.8 × 10$^6$, and 3.7 × 10$^6$, are also provided in Supplementary Note 6. 

The sub-shot-noise characteristics of the twin beams can be degraded by optical loss. In particular, the unbalanced loss applied to one of the twin beams leads to rapid degradation of the NRF. In contrast, when both beams experience the same loss, the degradation is mitigated as the NRF decreases linearly with the loss. Therefore, when loss occurs in one arm due to sample interrogation, it is essential to intentionally introduce a corresponding loss in the other arm to suppress NRF degradation. The theoretical and experimental discussions on how signal and idler losses affect the NRF are described in Supplementary Note 7.

\subsection*{MIP measurements of water}
Next, we perform MIP detection by irradiating water with an intensity-modulated QCL (4.64 $\mathrm{\mu m}$, 2153 cm$^{-1}$) as the pump and probing the induced response using a 1.55 $\mathrm{\mu m}$ coherent beam. The water absorbance at a pump wavenumber of 2153 cm$^{-1}$ is 0.97, as measured with a commercial FTIR spectrometer. In the present experiment, irradiating the QCL onto water caused the probe beam pattern in the Fourier plane to change from Gaussian-like to ring-like profiles (see Supplementary Notes 8 and 9). By placing an aperture in the Fourier plane, this beam-pattern variation is converted into an intensity change, which is measured as the MIP signal. The QCL intensity-modulation frequency is set to 2 MHz. At this frequency, the diffusion term in the heat-conduction equation can be neglected, while measurements can be performed without being limited by the classical noise at around 1 MHz of the probe beam, thereby enabling shot-noise-limited detection, as discussed below.

\begin{figure}[htbp]
\centering
\includegraphics[width=1\linewidth]{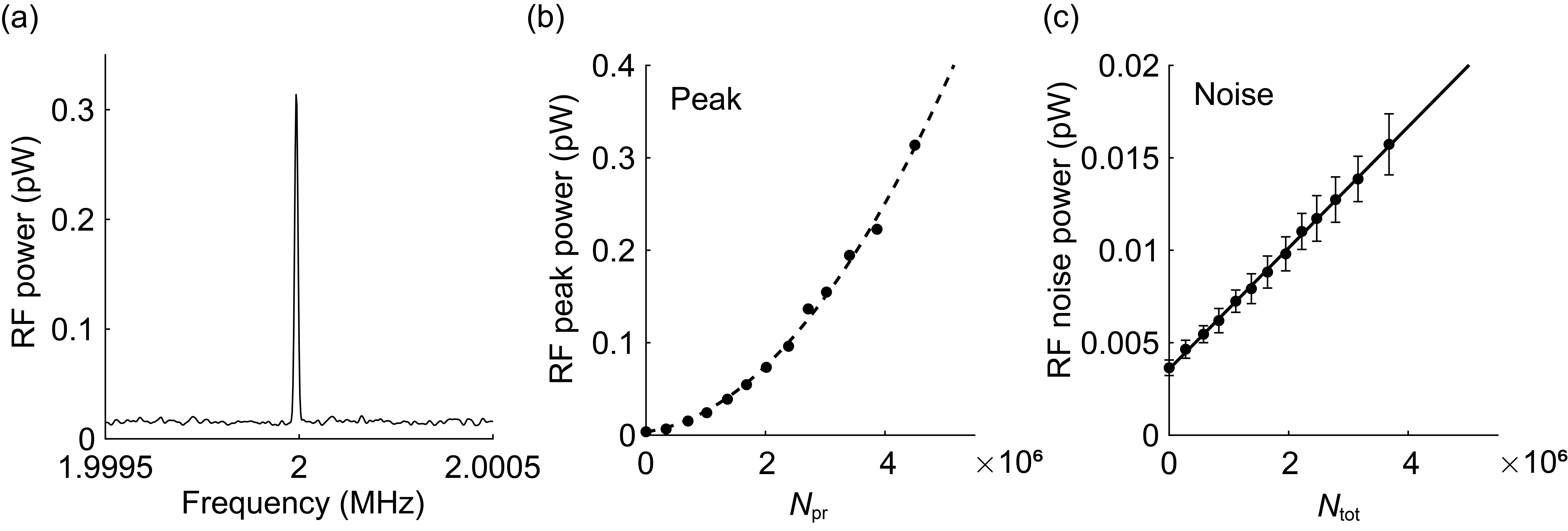}
\caption{MIP measurement of water. (a) MIP signal (RF power) measured using an RFSA in full-span mode. (b) The peak of the MIP signal as a function of $N_{\mathrm{pr}}$. The dashed line represents the quadratic fit. (c) The noise floor as a function of $N_{\mathrm{tot}}$. The solid line represents the linear fit. The y-axis error bars represent the standard deviation of the RF noise power measured over the frequency range from 1.9990 to 1.9998 MHz.}
\label{fig:MIP_shot}
\end{figure}

Figure~\ref{fig:MIP_shot}(a) shows the MIP signal obtained using the coherent beam as the probe. The probe beam, after experiencing the MIP-induced modulation, is detected by one port of the BPD. The detected signal is measured using an RFSA in full-span mode, with a center frequency of 2 MHz, a resolution bandwidth of 10 Hz, and a video bandwidth of 1 Hz. Modulation of the pump beam at 2 MHz produces an RF peak at the same frequency in the detected probe signal. Figures~\ref{fig:MIP_shot}(b) and (c) show the peak power of the MIP signal and the noise floor plotted as functions of the probe mean photon number per pulse $N_{\mathrm{pr}}$ and the total mean photon number per pulse $N_{\mathrm{tot}}$, respectively. Both $N_{\mathrm{pr}}$ and $N_{\mathrm{tot}}$ are calculated from the optical power measured before the BPF. The calculation of $N_{\mathrm{pr}}$ accounts for the BPF transmittance and optical losses at the sample position, whereas that of $N_{\mathrm{tot}}$ additionally includes residual reflection from the BPF and the ATT transmittance in the reference arm, as well as aperture losses in the probe arm. The corresponding dashed and solid lines represent the quadratic and linear fitting curves, respectively. The MIP-induced detector voltage is proportional to $N_{\mathrm{pr}}$. Since the RFSA measures the corresponding electrical power, which is proportional to the square of the voltage, the RF peak power scales quadratically with $N_{\mathrm{pr}}$. In contrast, the shot-noise power scales linearly with $N_{\mathrm{tot}}$. Consequently, the RF power SNR increases linearly with $N_{\mathrm{tot}}$. These results demonstrate that, by using a shot-noise-limited probe source and appropriately selecting the RF detection frequency, point-scanning MIP measurements with a shot-noise-limited SNR can be achieved.

\subsection*{Sub-shot-noise MIP}
\begin{figure}[htbp]
\centering
\includegraphics[width=0.5\linewidth]{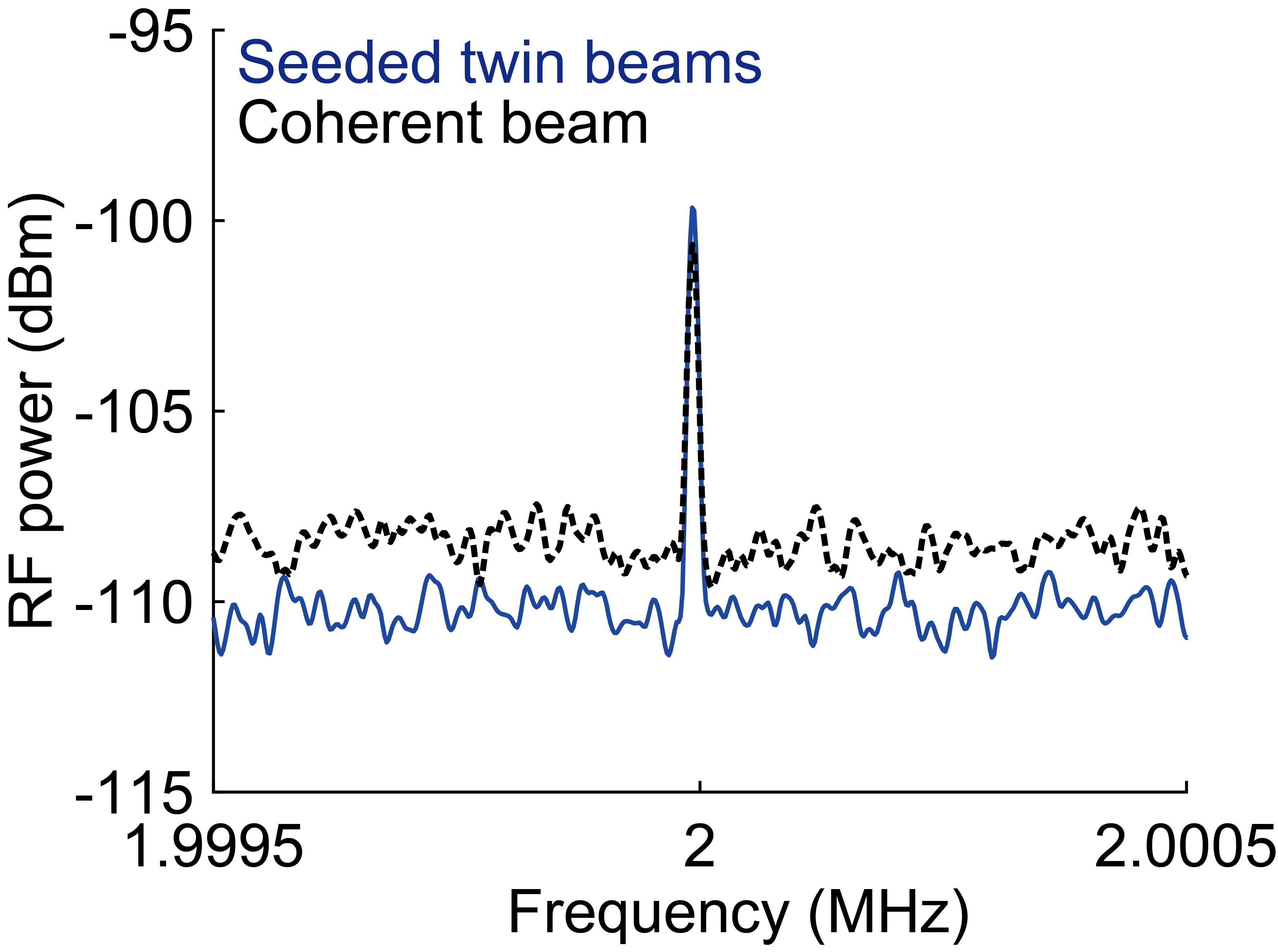}
\caption{MIP measurement with twin beams. Blue: RF spectra measured using twin beams with $N_{\mathrm{tot}}$ and $N_{\mathrm{pr}}$ of 3.5x10$^6$  and 2.6x10$^6$, respectively. Dashed black: RF spectra measured using a coherent beam with $N_{\mathrm{tot}}$ and $N_{\mathrm{pr}}$ of 3.4x10$^6$  and 2.1x10$^6$, respectively.}
\label{fig:RFspectra}
\end{figure}

We switch from MIP measurements with a coherent beam to seeded twin beams with balanced detection. In the twin beam measurements, the signal beam is detected at one port of the BPD after interacting with the sample, while the idler beam is detected at the other port. For comparison, balanced detection using a coherent beam is also carried out by replacing the BPF with a pellicle BS. Both $N_{\mathrm{tot}}$ and $N_{\mathrm{pr}}$ are calculated from the total optical power measured before the BPF for the twin beams or before the BS for the coherent beams, using $r_0$ and the reflection/transmission splitting ratios of the corresponding component. $N_{\mathrm{pr}}$ accounts for optical losses at the sample position in the probe arm, whereas $N_{\mathrm{tot}}$ additionally includes aperture losses in the probe arm and attenuation by the ATT in the reference arm. Figure~\ref{fig:RFspectra} compares the measured MIP signals obtained using a coherent beam and twin beams at $r_0=$2.0. The measurements are performed under conditions in which $N_{\mathrm{tot}}$ of the coherent and twin beams are comparable. In the twin-beam case, the noise floor is reduced by approximately 2 dB compared with the coherent-beam case, demonstrating that twin beams reduce the noise floor in balanced-detection MIP measurements. Note that, in the present experiment, $N_{\mathrm{pr}}$ is slightly higher in the twin-beam configuration than in the BS configuration under equal $N_{\mathrm{tot}}$ conditions, owing to the reflection/transmission imbalance of the pellicle BS and the signal/idler power imbalance of the twin beams. 

\begin{figure}[htbp]
\centering
\includegraphics[width=0.8\linewidth]{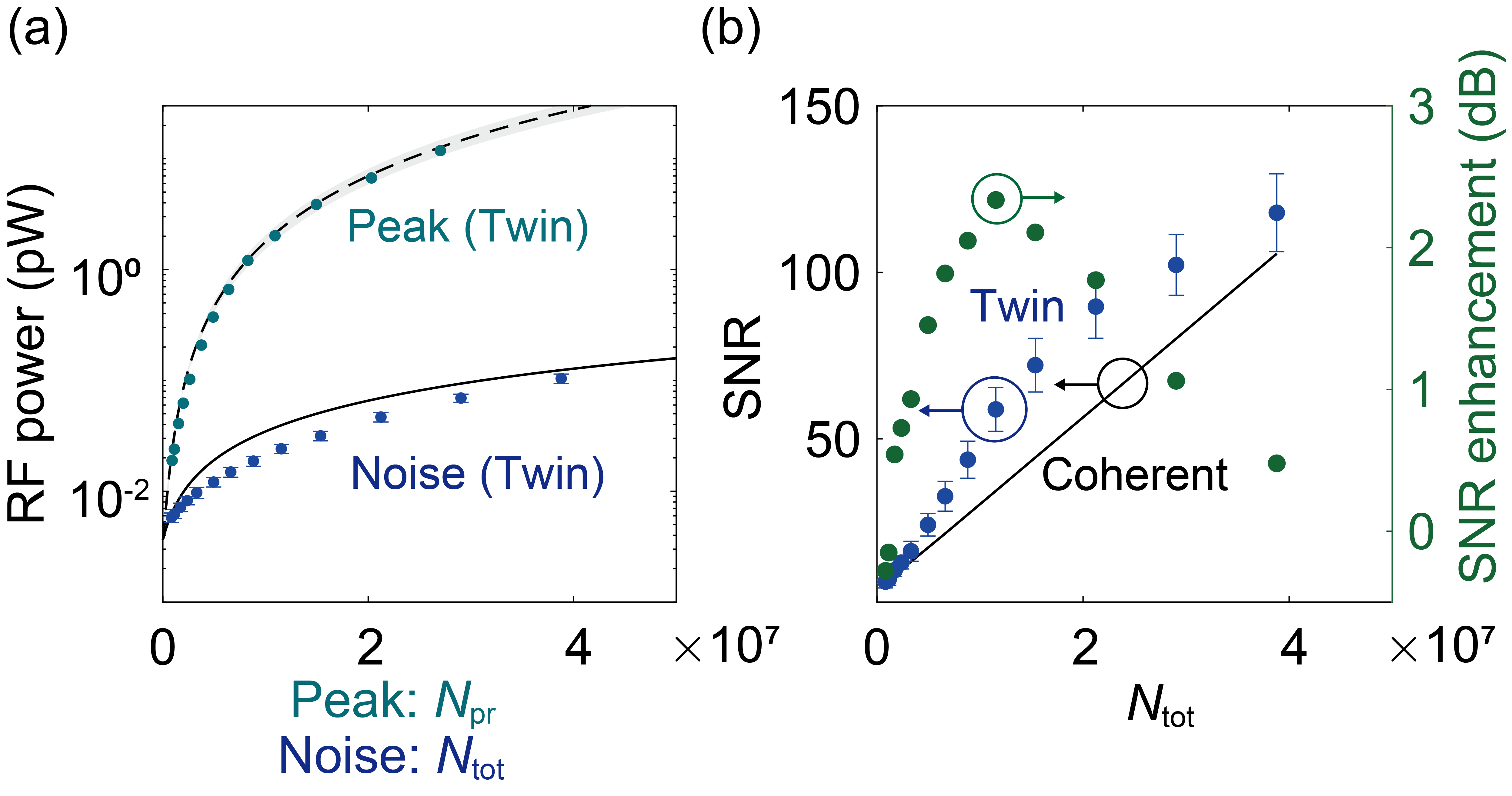}
\caption{Sub-shot-noise MIP measurement. (a) RF peak (blue-green) and noise (blue) measured using twin beams with $N_{\mathrm{0}}$ of 7.5x10$^5$ as functions of $N_{\mathrm{pr}}$ and $N_{\mathrm{tot}}$, respectively. The y-axis error bars for the noise represent the standard deviation of the RF noise power measured over the frequency range from 1.9990 to 1.9998 MHz. The dashed and solid black lines are the corresponding RF-peak and -noise power fitting curves of balanced coherent beams, respectively. The shaded gray areas represent the fitting errors for the dashed and solid black lines. (b) The corresponding SNR of the twin-beam measurements (blue) as a function of $N_{\mathrm{tot}}$, compared with the coherent-beam case (black line). The y-axis error bars indicate the standard deviation of the SNR calculated from the standard deviations of the noise using the error propagation law. The green plots show the SNR enhancement achieved using twin beams, calculated as the ratio of the SNR obtained with twin beams to that obtained with coherent beams.}
\label{fig:subshotnoise}
\end{figure}

To quantitatively evaluate the noise reduction of the MIP signal achieved with twin beams and the corresponding improvement in the SNR, we vary $r_0$ to control the mean photon numbers of the twin beams. We then measure the noise level as a function of $N_{\mathrm{tot}}$ and the peak power as a function of $N_{\mathrm{pr}}$, and compare the results with those obtained using a coherent beam. Figure~\ref{fig:subshotnoise}(a) shows the measurement results for twin beams with $N_{\mathrm{0}}$ of 7.5×10$^5$. In the coherent-beam case, the peak power and noise are fitted to quadratic and linear functions, respectively, as detailed in Supplementary Note 10. The peak power of the twin beams tends to follow the fitting curve of the coherent-beam probe, whereas the noise exhibits sub-shot-noise behavior, falling below the coherent-beam fitting curve at certain $N_{\mathrm{tot}}$. From the noise levels of the coherent and twin beams obtained in Fig. ~\ref{fig:subshotnoise}(a), for example, the NRF with detector noise subtraction is -$2.4$ dB at $N_{\mathrm{tot}}$ of 1.2x10$^7$. 

We further compare the SNR of the MIP signal (defined as the ratio of MIP peak to noise powers) for the coherent- and twin-beam cases in Fig. ~\ref{fig:subshotnoise}(b). The detector noise level is subtracted for the SNR calculation. The black line represents the result for the coherent beam with balanced detection, obtained from the fitting curves in Fig. ~\ref{fig:subshotnoise}(a). The twin-beam case shows a clear enhancement over the coherent-beam SNR curve. The SNR enhancement of 2.3 dB is obtained at $N_{\mathrm{tot}}$ of 1.2×10$^7$. 

\section{Discussion}
\subsection*{Twin beams}
This study employed pulsed twin beams generated from a seeded OPA in a PPLN waveguide as the quantum light source for MIP measurements. However, twin beams produced via alternative generation schemes can also be utilized. For instance, twin beams can be produced via $\chi^{(3)}$ nonlinearities, such as FWM in atom vapor cells~\cite{Slusher1985}, in optical fibers~\cite{Shelby1986}, or in microresonator~\cite{Dutt2015}. Indeed, in quantum-enhanced photothermal microscopy in the visible regime~\cite{Fu2026}, Fu et al. employed CW twin beams generated by FWM in a Rb vapor cell. The limited gain bandwidth of near-resonant Rb-vapor FWM favors CW operation or sufficiently long pulses~\cite{Agha2011}. Our 216.7-MHz source can also be regarded as a quasi-CW probe for our MIP demonstrations, as its repetition rate is much higher than the MIP modulation frequency. Nevertheless, its pulsed operation with high peak power enables efficient nonlinear optical interactions. This opens possibilities such as generating the MIR pump for MIP via difference-frequency generation (DFG) from the fundamental pulses, or applying portions of the fundamental pulse (or derived pulses) to nonlinear spectroscopy techniques on the sample~\cite{Hashimoto2019,Sinjab2020}. In this way, the intrinsic advantages of pulsed operation allow the integration of additional modalities into sub-shot-noise MIP measurements.

The MIP probe used in this study operates in the 1.5-µm band, where the ready availability of fiber components and waveguides facilitates system simplification~\cite{Zehavi2026}. For bio-imaging of cells or tissues, shorter-wavelength twin-beam or squeezed-light probes near 800 or 1064 nm may be more advantageous owing to lower background water absorption and the potential for higher spatial resolution.

\subsection*{Balanced detection}
In our approach, one of the twin beams serves as the MIP probe, and balanced detection is required to exploit the intensity correlations between the beams. Within this framework, shot-noise reduction and the associated improvement in SNR can be achieved compared to a coherent beam. However, when compared at the same $N_{\mathrm{pr}}$ as a single-port detection scheme using a coherent probe, the SNR is reduced by around a factor of two. Therefore, a noise reduction exceeding 3 dB is required to achieve a net quantum advantage over single-port coherent detection~\cite{Xu2022}. Alternatively, using both the signal and idler beams as MIP probes and positioning their foci before and after the thermal lens, respectively, could induce MIP modulations of opposite signs~\cite{Markus2012}. In this configuration, balanced detection of the twin beams would double the MIP signal. Another approach may employ amplitude-squeezed light generated via the displacement of a squeezed vacuum~\cite{deAndrade2020,Akatev2026}, Kerr squeezing~\cite{Herman2025}, or degenerate-seeded OPA~\cite{Casacio2021}, as recently demonstrated in spectroscopy applications. The amplitude-squeezed light enables sub-shot-noise MIP measurements in a single-port detection configuration.

\subsection*{MIP discussions}
In the present proof-of-principle experiment, a single-frequency QCL was used as the MIR pump to irradiate water, and the resulting MIP effect was probed by one of the twin beams. Further investigations are required to establish applicability to practical MIP spectroscopy. By employing a wavelength-tunable MIR source and sweeping the wavelength, one can verify that the measured MIP signal reflects the sample’s MIR absorption spectrum, thereby demonstrating spectroscopic capability. In this context, as noted above, MIR radiation can be generated via DFG from the 1.5 $\mathrm{\mu m}$ fundamental pulses. 

As discussed in Supplementary Note 8, the MIP phase shift in the present experiment decreases with the modulation frequency. We therefore selected a modulation frequency of 2 MHz within the 1–10 MHz range as a compromise between maintaining the sub-shot-noise performance and retaining a sufficiently strong MIP signal. In contrast, conventional point-scanning MIP microspectroscopy typically employs modulation frequencies on the order of 10–100 kHz~\cite{Yin2023}. Given that noise reduction using seeded pulsed twin beams and balanced detection has previously been demonstrated at an analysis frequency of around 200 kHz~\cite{Liu2018, Zehavi2026}, we anticipate that sub-shot-noise MIP detection should be feasible over a frequency range extending from hundreds of kHz to several MHz. Furthermore, operation at MHz modulation frequencies may enable MIP measurements in an energy-deposition regime, providing an alternative measurement strategy~\cite{Yin2025}.

\subsection*{Signal and the loss}
The MIR pump irradiation changes the sample’s refractive index, thereby altering the probe beam’s wavefront and transforming its Fourier-plane intensity profile from Gaussian-like to ring-like, as shown in Supplementary Note 9. To convert the change in spatial distribution into a modulation of the detected probe power, we progressively closed the iris at the Fourier plane, thereby varying the aperture transmittance. The ring-like beam profile causes the integrated transmitted power to vary non-monotonically with aperture size, resulting in multiple local maxima, with its largest observed value at an aperture transmittance of 0.42, as shown in Supplementary Note 9. For the present proof-of-principle demonstration, we set the aperture transmittance to 0.81 to retain sufficient signal-arm transmission for observing sub-shot-noise intensity-difference correlations. Reducing the transmittance from 0.81 to 0.42 is expected to increase the MIP RF peak power by a factor of approximately 4.9. Based on the corresponding total detected optical powers, the shot-noise power is estimated to decrease to approximately 0.48 of its value at 0.81. These changes would yield an approximately tenfold improvement in the RF-power SNR relative to a coherent-beam shot-noise reference. For the twin beams, however, the additional signal-arm loss also degrades the noise reduction. We predict an NRF of approximately -1 dB at a transmittance of 0.42 for $r_0$=2. 

\section{Conclusion}
In conclusion, we demonstrated a proof of principle for sub-shot-noise MIP measurements by inducing a photothermal effect in a water sample using a 2 MHz intensity-modulated MIR source and probing it with telecom-band pulsed twin beams generated via seeded OPA in a PPLN waveguide. We achieved a noise reduction of more than 5 dB with twin-beam intensity difference squeezing after subtracting detector noise, compared with shot-noise-limited coherent beams. When combined with MIP measurements, twin-beam detection reduced the noise floor and improved the SNR by approximately 2 dB. In the present demonstration, the measurement was limited to thermal-lens measurements using a single-frequency MIR pump. In future work, we will sweep the pump frequency to extract and quantitatively evaluate the absorption spectra of target samples, and investigate the technique's applicability to microscopic measurements by preparing microscopic-scale samples. Further advancing this method is expected to enable sensitive MIP measurements of fragile samples susceptible to photodamage.

\section{Methods}
The detailed schematic of the system is illustrated in Supplementary Note 1.
\subsection*{OPA pump generation}
The femtosecond oscillator, operating at a central wavelength of 1554.1 nm with a repetition rate of 216.67 MHz (MENHIR-1550, Menhir Photonics), was used as the fundamental source. The output beam was coupled into a polarization-maintaining (PM) fiber, and 10\% of the power was tapped using a 10:90 PM fiber coupler. The beam was spectrally filtered by an FBG (O/E Land) centered at 1560 nm with a bandwidth of 0.2 nm, and the reflected light was extracted using an optical circulator. The filtered beam was subsequently amplified by a two-stage fiber amplifier consisting of an Er-doped fiber amplifier (EDFA100P, Thorlabs) followed by a custom-built Er/Yb co-doped double-clad fiber amplifier. The amplified beam was launched into free space via a fiber collimator (F220APC-1550, Thorlabs). The polarization was adjusted using a quarter-wave plate (QWP), and unwanted spectral components, including amplified spontaneous emission (ASE), were suppressed using a 1560-nm BPF with a bandwidth of 12 nm (FBH1560-12, Thorlabs). The beam power was controlled using a half-wave plate (HWP) in combination with a polarizing beamsplitter (PBS), while the polarization angle after the PBS was further tuned using an additional HWP. The beam was then focused into a 50-mm-long PPLN crystal with a poling period of 19.36 µm (SHNIR-SA-50, HCPhotonics) using an $f =$ 100 mm lens to generate second-harmonic radiation. The crystal temperature was adjusted at 115.9 $^{o}$C using an oven. The SHG beam was collimated by another $f =$ 100 mm lens, and the residual fundamental beam was removed using a short-pass filter (SPF) with a cutoff wavelength of 950 nm (FESH0950, Thorlabs).

\subsection*{OPA seed preparation}
90\% of the fundamental beam from the 10:90 PM fiber coupler was directed to a custom-built spectral filtering unit comprising a fiber collimator(F810APC-1550, Thorlabs), a HWP, a diffraction grating with a groove density of 1200 lines/mm (GR25-1208, Thorlabs), a concave mirror, an adjustable slit, and an end mirror. The beam was launched into free space using the collimator, and its polarization was adjusted by the HWP. The beam was then dispersed by the grating and focused by the concave mirror. In the Fourier plane, the spectrum of the fundamental beam was filtered using the slit; the central wavelength and bandwidth were controlled by adjusting the position and width of the slit, respectively. The central wavelength was detuned from the degenerate wavelength for the OPA discussed later. The filtered beam was reflected back along the incident path, coupled into a fiber via an optical circulator, and subsequently re-collimated into free space using a fiber collimator (F220APC-1550, Thorlabs). The output beam was directed into a delay line consisting of a HWP, a PBS, a QWP, and a mirror. 

\subsection*{Twin-beam generation}
The SHG beam and the spectrally filtered fundamental beam described above were used as the pump and seed, respectively, for nondegenerate OPA to generate twin beams. The pump power was controlled by the combination of a HWP and a PBS and measured using a thermal power meter (S401C, Thorlabs). The seed beam divergence was slightly changed using an additional $f=$ 750 mm lens. After their linear polarization angles were independently adjusted using HWPs, the beams were combined using a dichroic mirror (DMSP1000, Thorlabs) with a cutoff wavelength of 1000 nm. The combined beams were focused into a 40-mm-long PPLN waveguide with a poling period of 18.2 µm (WH-0780-000-A-C-C-TEC-K02, NTT Innovative Devices) using an achromatic lens with a focal length of 10 mm (AC080-010-B-ML, Thorlabs). The waveguide temperature was adjusted at 37.2 $^{o}$C using a Peltier controlled by a temperature controller (C11330-01, Hamamatsu). The phase-insensitive OPA process within the waveguide generates signal and idler beams (i.e., seeded twin beams), symmetrically detuned from the degenerate frequency.  The generated twin beams were collimated by an $f=$ 8 mm aspheric lens (C240TMD-C, Thorlabs) and transmitted through a long-pass filter (LPF) with a cut-on wavelength of 1300 or 1400 nm (FELH1300 or FELH1400, Thorlabs) to remove the residual pump. They were subsequently directed onto a slightly tilted 1550-nm BPF with a bandwidth of 12 nm (FBH1550-12, Thorlabs), which spatially separates the signal and idler beams. The twin-beam power was measured before the BPF using a Ge power meter (S122C, Thorlabs). The transmitted beam (signal) was used as the probe for the photothermal measurements described below, while the reflected beam (idler) served as the reference. For comparison with the balanced detection of coherent beam cases, a pellicle BS (BP145B3, Thorlabs) was used.

\subsection*{MIP measurements}
A CW distributed-feedback QCL (DFB-QCL; QD4500CM1-AA309, Thorlabs) with a central wavenumber of 2153 cm$^{-1}$ (4.64 µm) was employed as the pump source for MIP measurements. The MIR beam was externally modulated at 2 MHz using an AOM (GEM-60-3-2000/5000/2mm, Brimrose) driven by a sinusoidal signal. The linear polarization angle was adjusted using a HWP before the AOM. The first-order diffracted beam was rejected, while the zeroth-order beam was coupled into an InF$_{3}$ single-mode fiber(P3-32F-FC-1,
Thorlabs), thereby imprinting a 2-MHz intensity modulation onto the MIR beam. The fiber output was collimated using a collimator (F028APC-3450, Thorlabs) and transmitted through a Ge window, where a weakly reflected component was detected by an InAsSb photodetector (PDA07P2, Thorlabs) to monitor the modulation of the MIR intensity. The transmitted MIR beam was subsequently focused into a water sample using a ZnSe lens with a focal length of 15 mm. The average MIR power under 2-MHz modulation, measured by a thermal power meter (S401C, Thorlabs) before the focusing lens, was around 15 mW. The sample was contained in a liquid cell (PIKE Technologies) consisting of two 3-mm-thick CaF$_{2}$ windows separated by a 50-µm-thick Teflon spacer. The probe beam, whose linear polarization was adjusted by a HWP, was focused into the same interaction region within the sample cell using a lens with a focal length of 75 mm. The pump beam was incident normal to the sample cell, whereas the probe beam was introduced at an oblique angle. After interaction with the sample, the probe beam was recollimated by a second lens with a focal length of 75 mm. The recollimated beam passed through an iris positioned at the Fourier plane to transfer the beam-pattern variation induced by the MIP process to the probe-beam intensity modulation. 

\subsection*{Detection and Digitization}
The probe (signal) beam transmitted through the iris was focused onto one input port of a 10-MHz BPD (fabricated by Graviton Inc.) using an $f=$ 50 mm lens. The BPD consists of two high-quantum-efficiency InGaAs photodiodes (FD500N-1550, Fermionics). The reference (idler) beam was similarly focused onto the other input port of the BPD using another an $f=$ 50 mm lens. The reference beam power was intentionally attenuated using a HWP and a PBS to match the optical loss in the probe arm, thereby preserving the noise reduction achieved by the balanced detection. The HWP rotation angle was adjusted to minimize the noise floor in the RF spectrum measured with the twin beams. The differential output voltage from the BPD was low-pass filtered at DC–11 MHz with a low-pass filter (BLP-10.7+, MiniCircuits), amplified by 10 dB using a low-noise voltage amplifier (DHPVA-101, FEMTO), and analyzed with an RFSA (N9938A, Keysight Technologies). For zero-span measurements, the center frequency, resolution bandwidth, and video bandwidth were set to 5 MHz, 1 MHz, and 1 kHz, respectively. For full-span measurements, the corresponding parameters were set to 2 MHz, 10 Hz, and 1 Hz, respectively. The fundamental, OPA pump, OPA seed, and twin beam spectra were evaluated using an OSA (AQ6374E, Yokogawa). 

\begin{backmatter}

\bmsection{Funding}
This work is supported by JST PRESTO (JPMJPR2457, K.H.), JST FOREST (JPMJFR236C, T.I.), JSPS KAKENHI (25H01386,23H00273, T.I.), RIKEN TRIP initiative (HIKARI-COOL Tokyo) (T.I.). 

\bmsection{Acknowledgment}
The authors acknowledge Dr. Zicong Xu for giving us valuable comments on this manuscript. K.H. acknowledges Prof. Maria Chekhova for valuable guidance on twin beams. K.H. also acknowledges Associate Prof. Yujiro Eto for his valuable advice on quantum sources and their detection. K.H. also acknowledges Dr. Keiichiro Toda and Mr. Genki Ishigane for their valuable advice on MIP measurements. The authors acknowledge Prof. Akira Furusawa and Associate Prof. Mamoru Endo for their support in developing the BPD, and Prof. Naoya Kuse for his support in developing a laser used in the initial experiments. The authors thank the Physics Department Workshop at The University of Tokyo for making custom-made optical mounts and enclosures. The authors also thank Omron Sentech for letting us use their demo version of the InGaAs camera.

\bmsection{Disclosures}
K.H. and T.I. are named inventors on a patent application related to the sub-shot-noise MIP measurement system. (P) 
\bmsection{Data Availability Statement}
The data provided in the manuscript are available from the corresponding author upon reasonable request.
\bmsection{Code Availability Statement}
The codes used to analyze the data of this study are available from the corresponding author upon reasonable request.

\bmsection{Author Contributions}
K.H. conceived the project. K.H. designed and built the optical setup in consultation with T.I. K.H. acquired and analyzed experimental data, and developed simulation codes. K.H. and T.I. discussed the results. K.H. supervised the work. K.H. wrote the manuscript with input from T.I. 

\bmsection{Supplemental document}
See Supplement 1 for supporting content.
\end{backmatter}

\clearpage
\bibliography{references}

@article{Ishigane2023,
  author    = {G. Ishigane and K. Toda and M. Tamamitsu and H. Shimada and V. R. Badarla and T. Ideguchi},
  title     = {Label-free mid-infrared photothermal live-cell imaging beyond video rate},
  journal   = {Light Sci. Appl.},
  year      = {2023},
  volume    = {12},
  pages     = {174},
}

@article{Yin2023,
  author    = {J. Yin and M. Zhang and Y. Tan and Z. Guo and H. He and L. Lan and J.-X. Cheng},
  title     = {Video-rate mid-infrared photothermal imaging by single-pulse photothermal detection per pixel},
  journal   = {Sci. Adv.},
  year      = {2023},
  volume    = {9},
  pages     = {eadg8814},
}

@article{Xu2022,
  author    = {Z. Xu and K. Oguchi and Y. Taguchi and Y. Sano and Y. Miyawaki and D. Cheon and K. Katoh and Y. Ozeki},
  title     = {Stimulated Raman scattering spectroscopy with quantum-enhanced balanced detection},
  journal   = {Opt. Express},
  year      = {2022},
  volume    = {30},
  pages     = {18589-18598},
}

@article{deAndrade2020,
  author    = {R. B. de Andrade and H. Kerdoncuff and K. Berg-Sørensen and T. Gehring and M. Lassen and U. L. Andersen},
  title     = {Quantum-enhanced continuous-wave stimulated Raman scattering spectroscopy},
  journal   = {Optica},
  year      = {2020},
  volume    = {7},
  pages     = {470-475},
}

@article{Casacio2021,
  author    = {C. A. Casacio and L. S. Madsen and A. Terrasson and M. Waleed and K. Barnscheidt and B. Hage and M. A. Taylor and W. P. Bowen},
  title     = {Quantum-enhanced nonlinear microscopy},
  journal   = {Nature},
  year      = {2021},
  volume    = {594},
  pages     = {201-206},
}

@article{Akatev2026,
  author    = {D. Akatev and Y. Meng and J. Brewer and M. V. Chekhova and U. L. Andersen and M. Lassen},
  title     = {Broadly tunable quantum-enhanced Raman microscopy for advancing bioimaging},
  journal   = {Optica Quantum},
  year      = {2026},
  volume    = {4},
  pages     = {108-113},
}

@article{Wan2025,
  author    = {Z. Wan and Y. Chen and X. Zhang and M. Yan and H. Zeng},
  title     = {Quantum correlation-enhanced dual-comb spectroscopy},
  journal   = {Light Sci. Appl.},
  year      = {2025},
  volume    = {14},
  pages     = {257},
}

@article{Herman2025,
  author    = {D. I. Herman and M. Walsh and M. K. Kreider and N. Lordi and E. J. Tsao and A. J. Lind and M. Heyrich and J. Combes and J. Genest and S. A. Diddams},
  title     = {Squeezed dual-comb spectroscopy},
  journal   = {Science},
  year      = {2025},
  volume    = {387},
  pages     = {653-658},
}

@misc{Fu2026,
  author    = {P. Fu and X. Liu and S. Wang and N. Li and C. Xu and H. Cai and H. Hu and V. V. Yakovlev and X. Liu and S.-Y. Zhu and X. Xu and D. Zhang and D.-W. Wang},
  title     = {Quantum Squeezing Enhanced Photothermal Microscopy},
  journal   = {arXiv},
  eprint    = {2601.20632},
archivePrefix = {arXiv},
  doi           = {10.48550/arXiv.2601.20632},
    year      = {2026},
}

@article{Xia2022,
  author    = {Q. Xia and J. Yin and Z. Guo and J.-X. Cheng},
  title     = {Mid-Infrared Photothermal Microscopy: Principle, Instrumentation, and Applications},
  journal   = {The Journal of Physical Chemistry B},
  year      = {2022},
   volume    = {126},
  pages     = {8597},
}

@article{Andersen2016,
  author    = {U. L Andersen and T. Gehring and C. Marquardt and G. Leuchs},
  title     = {30 years of squeezed light generation},
  journal   = {Phys. Scr.},
  year      = {2016},
   volume    = {91},
  pages     = {053001},
}

@article{Lawrie2019,
  author    = {B. J. Lawrie and P. D. Lett and A. M. Marino and R. C. Pooser},
  title     = {Quantum Sensing with Squeezed Light},
  journal   = {ACS Photonics},
  year      = {2019},
   volume    = {6},
  pages     = {1307-1318},
}

@article{Chekhova2015,
  author    = {M.V.Chekhova and G.Leuchs and M.Żukowski},
  title     = {Bright squeezed vacuum: Entanglement of macroscopic light beams},
  journal   = {Opt. Commun.},
  year      = {2015},
   volume    = {337},
  pages     = {27-43},
}

@article{Meda2017,
  author    = {A. Meda and E. Losero and N. Samantaray and F. Scafirimuto and S. Pradyumna and A. Avella and I. Ruo-Berchera and M Genovese},
  title     = {Photon-number correlation for quantum
enhanced imaging and sensing},
  journal   = {J. Opt.},
  year      = {2017},
   volume    = {19},
  pages     = {094002},
}

@article{Heidmann1987,
  author    = {A. Heidmann and R. J. Horowicz and S. Reynaud and E. Giacobino and C. Fabre and G. Camy},
  title     = {Observation of Quantum Noise Reduction on Twin Laser Beams},
  journal   = {PRL},
  year      = {1987},
   volume    = {59},
  pages     = {2555},
}

@article{Boyer2008,
  author    = {V. Boyer and A. M. Marino and R. C. Pooser and P. D. Lett},
  title     = {Entangled Images from Four-Wave Mixing},
  journal   = {Science},
  year      = {2008},
  volume    = {321},
  pages     = {544-547},
}

@article{SabinesChesterking2019,
  author    = {J. Sabines-Chesterking and A. R. McMillan and P. A. Moreau and S. K. Joshi and S. Knauer and E. Johnston and J. G. Rarity and J. C. F. Matthews},
  title     = {Twin-beam sub-shot-noise raster-scanning microscope},
  journal   = {Optics Express},
  year      = {2019},
  volume    = {27},
  pages     = {30810-30818},
}

@article{Samantaray2017,
  author    = {N. Samantaray and I. Ruo-Berchera and A. Meda and M. Genovese},
  title     = {Realization of the first sub-shot-noise wide field microscope},
  journal   = {Light Sci. Appl.},
  year      = {2017},
  volume    = {6},
  pages     = {e17005},
}

@article{Brida2010,
  author    = {G. Brida and M. Genovese and I. Ruo-Berchera},
  title     = {Experimental realization of sub-shot-noise quantum imaging},
  journal   = {Nat. Photon.},
  year      = {2010},
  volume    = {4},
  pages     = {227-230},
}

@article{Li2021,
  author    = {F. Li and T. Li and M. O. Scully and G. S. Agarwal},
  title     = {Quantum Advantage with Seeded Squeezed Light for Absorption Measurement},
  journal   = {Phys. Rev. Applied},
  year      = {2021},
  volume    = {15},
  pages     = {044030},
}

@article{Moreau2017,
  author    = {P.-A. Moreau and J. Sabines-Chesterking and R. Whittaker and S. K. Joshi and P. M. Birchall and A. McMillan and J. G. Rarity and J. C. F. Matthews},
  title     = {Demonstrating an absolute quantum advantage in direct absorption measurement},
  journal   = {Sci. Rep.},
  year      = {2017},
  volume    = {7},
  pages     = {6256},
}

@article{Li2022,
  author    = {T. Li and F. Li and X. Liu and V. V. Yakovlev and G. S. Agarwal},
  title     = {Quantum-enhanced stimulated Brillouin scattering spectroscopy and imaging},
  journal   = {Optica},
  year      = {2022},
  volume    = {9},
  pages     = {959-964},
}

@article{Ather2023,
  author    = {H. Ather and H. An and H. Owens and S. Alajlouni and A. Shakouri and M. Hosseini},
  title     = {Quantum Sensing of Thermoreflectivity in Electronics},
  journal   = {Phys. Rev. Applied},
  year      = {2023},
  volume    = {19},
  pages     = {044040},
}

@article{Adamou2025,
  author    = {D. Adamou and L. Hirsch and T. Shields and S. Yoon and A. C. Dada and J. M. R. Weaver and D. Faccio and M. Peccianti and L. Caspani and M. Clerici},
  title     = {Quantum-enhanced time-domain spectroscopy},
  journal   = {Science Advances},
  year      = {2025},
  volume    = {11},
  pages     = {eadt2187},
}

@article{Matsuzaki2022,
  author    = {K. Matsuzaki and T. Tahara},
  title     = {Superresolution concentration measurement realized by sub-shot-noise absorption spectroscopy},
  journal   = {Nat. Commun.},
  year      = {2022},
  volume    = {13},
  pages     = {953},
}

@article{Whittaker2017,
  author    = {R. Whittaker and C. Erven and A. Neville and M. Berry and J. L. O’Brien and H. Cable and J. C. F. Matthews},
  title     = {Absorption spectroscopy at the ultimate quantum limit from single-photon states},
  journal   = {New J. Phys.},
  year      = {2017},
  volume    = {19},
  pages     = {023013},
}

@article{Mueller2020,
  author    = {J. D. Mueller and N. Samantaray and J. C. F. Matthews},
  title     = {A practical model of twin-beam experiments for sub-shot-noise absorption measurements},
  journal   = {Appl. Phys. Lett.},
  year      = {2020},
  volume    = {117},
  pages     = {034001},
}

@article{Agafonov2010,
  author    = {I. N. Agafonov and M. V. Chekhova and G. Leuchs},
  title     = {Two-color bright squeezed vacuum},
  journal   = {Physical Review A},
  year      = {2010},
  volume    = {82},
  pages     = {011801},
}

@article{Shelby1986,
  author    = {R. M. Shelby and M. D. Levenson and S. H. Perlmutter and R. G. DeVoe and D. F. Walls},
  title     = {Broad-Band Parametric Deamplification of Quantum Noise in an Optical Fiber},
  journal   = {PRL},
  year      = {1986},
  volume    = {57},
  pages     = {691},
}

@article{Slusher1985,
  author    = {R. E. Slusher and L. W. Hollberg and B. Yurke and J. C. Mertz and J. F. Valley},
  title     = {Observation of Squeezed States Generated by Four-Wave Mixing in an Optical Cavity},
  journal   = {PRL},
  year      = {1985},
  volume    = {55},
  pages     = {2409},
}

@article{Hashimoto2019,
  author    = {K. Hashimoto and V. R. Badarla and A. Kawai and T. Ideguchi},
  title     = {Complementary vibrational spectroscopy},
  journal   = {Nat. Commun.},
  year      = {2019},
  volume    = {10},
  pages     = {4411},
}

@article{Sinjab2020,
  author    = {F. Sinjab and K. Hashimoto and V. R. Badarla and J. Omachi and T. Ideguchi},
  title     = {Multimodal laser-scanning nonlinear optical microscope with a rapid broadband Fourier-transform coherent Raman modality},
  journal   = {Opt. Express},
  year      = {2020},
  volume    = {28},
  pages     = {20794-20807},
}

@article{Yin2025,
  author    = {J. Yin and C. Pfluegl and C. C. Teng and R. Bolarinho and G. Chen and X. Gong and D. Dong and D. Vakhshoori and J.-X. Cheng},
  title     = {Mid-Infrared Energy Deposition Spectroscopy},
  journal   = {PRL},
  year      = {2025},
  volume    = {134},
  pages     = {093804},
}

@article{Agha2011,
  author    = {I. H Agha and C. Giarmatzi and Q. Glorieux and T. Coudreau and P. Grangier and G. Messin},
  title     = {Time-resolved detection of relative-intensity squeezed nanosecond pulses in an 87Rb vapor},
  journal   = {New J. Phys.},
  year      = {2011},
  volume    = {13},
  pages     = {043030},
}

@article{Yin2025_2,
  author    = {J. Yin and P.-T. Lyu and R. Bolarinho and Y. Zhu and X. Ge and H. Ni and J.-X. Cheng},
  title     = {Vibrational photothermal imaging: theory, instrumentation, and applications},
  journal   = {Optica},
  year      = {2025},
  volume    = {12},
  pages     = {1367-1387},
}

@article{Harder2016,
  author    = {G. Harder and T. J. Bartley and A. E. Lita and S. W. Nam and T. Gerrits and C. Silberhorn},
  title     = {Single-Mode Parametric-Down-Conversion States with 50 Photons as a Source for Mesoscopic Quantum Optics},
  journal   = {PRL},
  year      = {2016},
  volume    = {116},
  pages     = {143601},
}

@article{Markus2012,
  author    = {M. Selmke and M. Braun and F. Cichos},
  title     = {Photothermal Single-Particle Microscopy: Detection of a Nanolens},
  journal   = {ACS Nano},
  year      = {2012},
  volume    = {6},
  pages     = {2741-2749},
}

@article{Ideguchi2025,
  author    = {T. Ideguchi},
  title     = {Mid-infrared photothermal microscopy},
  journal   = {JSAP Rev.},
  year      = {2025},
  volume    = {2025},
  pages     = {250209},
}

@article{Hirota2026,
  author    = {K. Hirota and T. Kashiwazaki and G. Ha and T. Yamashima and P. Jaturaphagorn and T. Suzuki and K. Takahashi and A. Kawasaki and A. Inoue and W. Asavanant and M. Endo and T. Umeki and A. Furusawa},
  title     = {Generation of 10-dB squeezed light from a broadband waveguide optical parametric amplifier with improved phase locking method},
  journal   = {Opt. Express},
  year      = {2026},
  volume    = {34},
  pages     = {7958-7966},
}

@article{Long1976,
  author    = {M. E. Long and R. L. Swofford and A. C. Albrecht},
  title     = {Thermal Lens Technique: A New Method of Absorption Spectroscopy},
  journal   = {Science},
  year      = {1976},
  volume    = {191},
  pages     = {183-185},
}

@article{Gaiduk2010,
  author    = {A. Gaiduk and M. Yorulmaz and P. V. Ruijgrok and M. Orrit},
  title     = {Room-Temperature Detection of a Single Molecule’s Absorption by Photothermal Contrast},
  journal   = {Science},
  year      = {2010},
  volume    = {330},
  pages     = {353-356},
}

@article{Boyer2002,
  author    = {D. Boyer and P. Tamarat and A. Maali and B. Lounis and M. Orrit},
  title     = {Photothermal Imaging of Nanometer-Sized Metal Particles Among Scatterers},
  journal   = {Science},
  year      = {2002},
  volume    = {297},
  pages     = {1160-1163},
}

@article{SoutoRibeiro1997,
  author    = {P. H. Souto Ribeiro and C. Schwob and A. Maître and C. Fabre},
  title     = {Sub-shot-noise high-sensitivity spectroscopy with optical parametric oscillator twin beams},
  journal   = {Opt. Lett.},
  year      = {1997},
  volume    = {22},
  pages     = {1893-1895},
}

@article{Jakeman1986,
  author    = {E. Jakeman and J.G. Rarity},
  title     = {The use of pair production processes to reduce quantum noise in transmission measurements},
  journal   = {Opt. Commun.},
  year      = {1986},
  volume    = {59},
  pages     = {219-223},
}

@article{Wu1991,
  author    = {J. Wu and T. Kitamori and T. Sawada},
  title     = {Ultramicrospectrochemical Analysis and Trace Determination
of Compounds Adsorbed on a Single Microparticulate Sample
by the Optical Beam Deflection Method},
  journal   = {Anal. Chem.},
  year      = {1991},
  volume    = {63},
  pages     = {217-219},
}

@article{Eto2007,
  author    = {Y. Eto and T. Tajima and Y. Zhang and T. Hirano},
  title     = {Observation of squeezed light at 1.535µm using a pulsed homodyne detector},
  journal   = {Opt. Lett.},
  year      = {2007},
  volume    = {32},
  pages     = {1698-1700},
}

@article{Dutt2015,
  author    = {A. Dutt and K. Luke and S. Manipatruni and A. L. Gaeta and P. Nussenzveig and M. Lipson},
  title     = {On-Chip Optical Squeezing},
  journal   = {Phys. Rev. Applied},
  year      = {2015},
  volume    = {3},
  pages     = {044005},
}

@article{Waclawek2019,
  author    = {J. P. Waclawek and C. Kristament and H. Moser and B. Lendl},
  title     = {Balanced-detection interferometric cavity-assisted photothermal spectroscopy},
  journal   = {Optics Express},
  year      = {2019},
  volume    = {27},
  pages     = {12183-12195},
}

@article{Bai2019,
  author    = {Y. Bai and D. Zhang and L. Lan and Y. Huang and K. Maize and A. Shakouri and J.-X. Cheng},
  title     = {Ultrafast chemical imaging by widefield photothermal sensing of infrared absorption},
  journal   = {Sci. Adv.},
  year      = {2019},
  volume    = {5},
  pages     = {eaav7127},
}

@article{Paiva2022,
  author    = {E. M. Paiva and F. M. Schmidt},
  title     = {Ultrafast Widefield Mid-Infrared Photothermal Heterodyne Imaging},
  journal   = {Anal. Chem.},
  year      = {2022},
  volume    = {94},
  pages     = {14242-14250},
}

@article{Zhang2021,
  author    = {Y. Zhang and C. Yurdakul and A. J. Devaux and L. Wang and X. G. Xu and J. H. Connor and M. S. Ünlü and J.-X. Cheng},
  title     = {Vibrational Spectroscopic Detection of a Single Virus by Mid-Infrared Photothermal Microscopy},
  journal   = {Analytical Chemistry},
  year      = {2021},
  volume    = {93},
  pages     = {4100-4107},
}

@article{Liu2018,
  author    = {Y. Liu and N. Huo and J. Li and X. Li},
  title     = {Long-distance distribution of the telecom band intensity difference squeezing generated in a fiber optical parametric amplifier},
  journal   = {Opt. Lett.},
  year      = {2018},
  volume    = {43},
  pages     = {5559-5562},
}

@article{Domínguez-Juárez2023,
  author    = {J. L. Domínguez-Juárez and R. Quintero-Torres and M. A. Cardoso-Duarte and M. A. Quiroz-Juárez and J. L. Aragón and J. Villatoro},
  title     = {Unveiling the properties of liquids via photothermal-induced diffraction patterns},
  journal   = {Commun. Phys.},
  year      = {2023},
  volume    = {6},
  pages     = {154},
}

@article{Zehavi2026,
  author    = {E. Zehavi and A. Bernstein and M. Natan and Y. London and O. Buch and A. Zadok and A. Pe’er},
  title     = {Bright seeded and quasi-continuous two-mode intensity squeezing in commercial fiber},
  journal   = {Opt. Express},
  year      = {2026},
  volume    = {34},
  pages     = {27749-27756},
}
\end{document}


\maketitle

\section*{Supplementary Note 1: Detailed schematic}
\begin{figure}[htbp]
\centering
\includegraphics[width=1\linewidth]{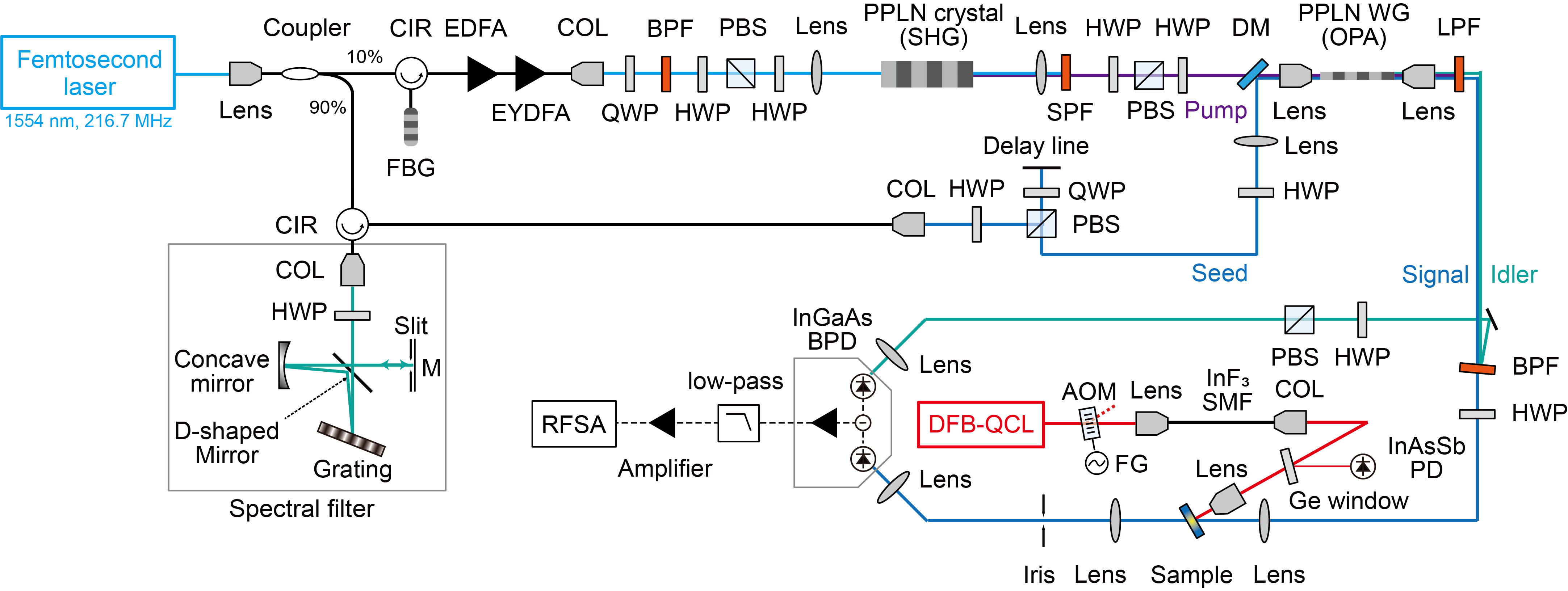}
\caption{Detailed schematic of the sub-shot-noise MIP measurement system using seeded twin beams. CIR: circulator, HWP: half-wave plate, COL: collimator, M: mirror, FBG: fiber-Bragg grating, EDFA: Er-doped fiber amplifier, EYDFA: Er/Yb co-doped fiber amplifier, QWP: quater-wave plate, BPF: band-pass filter, PBS: polarizing beamsplitter, PPLN: periodically poled lithium niobate, SPF: short-pass filter, DM: dichroic mirror, PPLN WG: PPLN waveguide, LPF: long-pass filter, DFB-QCL: distributed feedback quantum cascade laser, AOM: acousto-optic modulator, FG: functuion generator, SMF: single-mode fiber, InGaAs BPD: InGaAs balance photodetector, RFSA: RF spectrum analyzer.}
\label{fig:detailedschematic}
\end{figure}

The detailed schematic of the sub-shot-noise mid-infrared photothermal (MIP) measurement system is explained in Methods in the main text and illustrated in Figs.~\ref{fig:detailedschematic}.

\section*{Supplementary Note 2: Theory of seeded twin beams}
We theoretically model seeded twin beams and derive their noise-reduction factor (NRF). To model the twin beams, we approximate the signal and idler fields by a pair of modes, denoted by $a$ and $b$, respectively. We then individually introduce linear loss in each mode, derive the mean photon number in each detected mode, and calculate the variance of their photon-number difference.  

\subsection*{Two-mode Bogoliubov transformation}
The field evolution through the OPA is described by the two-mode Bogoliubov transformation~\cite{Chekhova2015}:
\begin{align}
a_1 &= \mu_0 a_0 + \nu_0^* b_0^\dagger, \\
b_1 &= \mu_0 b_0 + \nu_0^* a_0^\dagger,
\end{align}
where $\mu_0 = \cosh(r_0) e^{i \phi_{\mu_0}}, \nu_0 = \sinh(r_0) e^{i \phi_{\nu_0}},$ $r_0$ is the squeezing parameter, and $\phi_{\mu_0}, \phi_{\nu_0}$ are phases. The Bogoliubov coefficients satisfy $|\mu_0|^2-|\nu_0|^2=1$. The input creation and annihilation operators satisfy $[a_0, a_0^\dagger] = 1$ and $ [b_0, b_0^\dagger] = 1$, and the two input modes are assumed to be independent: $[a_0,b_0]=[a_0,b_0^\dagger]=0$. Each mode experiences linear loss after the waveguide:
\begin{align}
a_{2} &= t_{a} a_1 + r_{a} a_{\mathrm{vac}}, \\
b_{2} &= t_{b} b_1 + r_{b} b_{\mathrm{vac}},
\end{align}
where $t_a$ and $t_b$ are the complex amplitude transmittances,
and $r_a$ and $r_b$ are corresponding complex amplitude
reflectivities. They satisfy $|t_a|^2+|r_a|^2=1$ and $|t_b|^2+|r_b|^2=1$, and the corresponding intensity transmittances are $T_a=|t_a|^2$ and $ T_b=|t_b|^2$. The operators $a_{\mathrm{vac}}$ and $b_{\mathrm{vac}}$ represent independent vacuum modes and satisfy $[a_{\mathrm{vac}}, a_{\mathrm{vac}}^\dagger] = [b_{\mathrm{vac}}, b_{\mathrm{vac}}^\dagger] = 1, [a_{\mathrm{vac}},b_{\mathrm{vac}}]=[a_{\mathrm{vac}},b_{\mathrm{vac}}^\dagger]=0$. The vacuum modes are also assumed to be independent of the input modes: $[a_{\mathrm{vac}},a_0]=[a_{\mathrm{vac}},b_0]=[b_{\mathrm{vac}},a_0]=[b_{\mathrm{vac}},b_0]=0$, together with their Hermitian conjugates. 
The detected photon-number operators are defined as
\begin{align}
N_a=a_2^\dagger a_2,\qquad
N_b=b_2^\dagger b_2.
\end{align}
The total mean photon number is
\begin{align}
\langle N_{a}+N_{b} \rangle &= \langle N_{a} \rangle+\langle N_{b} \rangle.
\end{align}
The variance of the photon-number difference is
\begin{align}
\mathrm{Var}(N_{a} - N_{b}) =\mathrm{Var}(N_{a}) + \mathrm{Var}(N_{b}) - 2 \mathrm{Cov}(N_{a}, N_{b}),
\end{align}
where
$\mathrm{Var}(N_a)=
\langle N_a^2\rangle-\langle N_a\rangle^2, \mathrm{Var}(N_b)=
\langle N_b^2\rangle-\langle N_b\rangle^2, \mathrm{Cov}(N_a,N_b)=
\langle N_aN_b\rangle-\langle N_a\rangle\langle N_b\rangle,
$ and the symbol $\mathrm{Cov}$ denotes covariance.

\subsection*{Total mean photon numbers}
We assume that Mode $a$ is seeded by a coherent state
$|\alpha\rangle_{a_0}$, whereas Mode $b$ and the two loss ports
are in independent vacuum states. The total
input state is
\begin{align}
|\psi_{\mathrm{in}}\rangle
=|\alpha\rangle_{a_0}|0\rangle_{b_0}|0\rangle_{a_{\mathrm{vac}}}|0\rangle_{b_{\mathrm{vac}}}.
\end{align}
The coherent state satisfies
\begin{align}
a_0|\alpha\rangle_{a_0}
=
\alpha|\alpha\rangle_{a_0},
\qquad
N_0=|\alpha|^2,
\end{align}
where $N_0$ is the mean input photon number in the seeded mode.
The mean photon numbers in Modes $a$ and $b$ are given by
\begin{align}
\langle N_{a} \rangle &= T_{a}( N_0  \cosh^2 r_0 + \sinh^2 r_0),\\
\langle N_{b} \rangle &= T_{b}( N_0  + 1)  \sinh^2 r_0. 
\end{align}
The total mean photon number is
\begin{align}
\langle N_a+N_b\rangle &=N_0(T_a\cosh^2r_0+T_b\sinh^2r_0)+(T_a+T_b)\sinh^2r_0.
\end{align}
For a sufficiently strong seed satisfying $N_0(T_a\cosh^2r_0+T_b\sinh^2r_0)\gg(T_a+T_b)\sinh^2r_0$, the second term can be neglected:
\begin{align}
\langle N_a+N_b\rangle &\approx N_0(T_a\cosh^2r_0+T_b\sinh^2r_0).
\end{align}
In the absence of loss, $T_a=T_b=1$, $\langle N_{a} + N_{b} \rangle \approx N_0 \cosh 2r_0.$

\subsection*{Variance of photon-number difference}
The variance of the photon-number difference is 
\begin{align}
\mathrm{Var}(N_{a} - N_{b}) 
&= N_0 \Big[
2 T_{b}^2 \sinh^4 r_0
+ 2 T_{a}^2 \sinh^2 r_0 \cosh^2 r_0
- 4 T_{a} T_{b} \sinh^2 r_0 \cosh^2 r_0 \nonumber \\
&\quad + T_{b} \sinh^2 r_0
+ T_{a} \cosh^2 r_0
\Big] \nonumber \\
&\quad + T_{a}^2 \sinh^4 r_0
+ T_{b}^2 \sinh^4 r_0
- 2 T_{a} T_{b} \sinh^2 r_0 \cosh^2 r_0 \nonumber \\
&\quad + T_{a} \sinh^2 r_0
+ T_{b} \sinh^2 r_0.
\end{align}
For a sufficiently strong seed so that the terms independent of
$N_0$ are negligible, the variance is
\begin{align}
\mathrm{Var}(N_{a} - N_{b}) 
&\approx N_0 \Big[
2 T_{b}^2 \sinh^4 r_0
+ 2 T_{a}^2 \sinh^2 r_0 \cosh^2 r_0
- 4 T_{a} T_{b} \sinh^2 r_0 \cosh^2 r_0 \nonumber \\
&\quad + T_{b} \sinh^2 r_0
+ T_{a} \cosh^2 r_0 
\Big].
\end{align}
In the case where the loss is balanced, $T_a=T_b=T_1$, $\mathrm{Var}(N_{a} - N_{b}) \approx N_0 T_1 (\cosh 2r_0 - 2T_1 \sinh^2 r_0).$

\subsection*{NRF}
The shot-noise reference is defined using statistically independent
coherent beams with the same detected mean photon numbers as
Modes $a$ and $b$. The variance of the photon-number difference is,
\begin{align}
\mathrm{Var}_{\mathrm{shot}}(N_a-N_b)
&=
\mathrm{Var}_{\mathrm{shot}}(N_a)
+
\mathrm{Var}_{\mathrm{shot}}(N_b)
\nonumber\\
&=
\langle N_a\rangle+\langle N_b\rangle.
\end{align}
The NRF~\cite{Agafonov2010,Mueller2020} is 
\begin{align}
\mathrm{NRF} &= \frac{\mathrm{Var}(N_{a} - N_{b})}{\mathrm{Var}_{\mathrm{shot}}(N_{a} - N_{b})}\nonumber\\
&=\frac{\mathrm{Var}(N_{a} - N_{b})}{\langle N_{a} + N_{b} \rangle}.
\end{align}
For balanced loss, $T_a=T_b=T_1$, and in the strong-seed limit,
the NRF is
\begin{align}
\mathrm{NRF} \approx (1-T_1)+\frac{T_1}{\cosh 2r_0}.
\end{align}
The NRF depends linearly on the balanced intensity transmittance $T_1$. In the absence of loss, the NRF can be approximated as:
\begin{align}
\mathrm{NRF} \approx \frac{1}{\cosh 2r_0}.
\end{align}

\section*{Supplementary Note 3: Fundamental and SHG spectra}
\begin{figure}[htbp]
\centering
\includegraphics[width=1\linewidth]{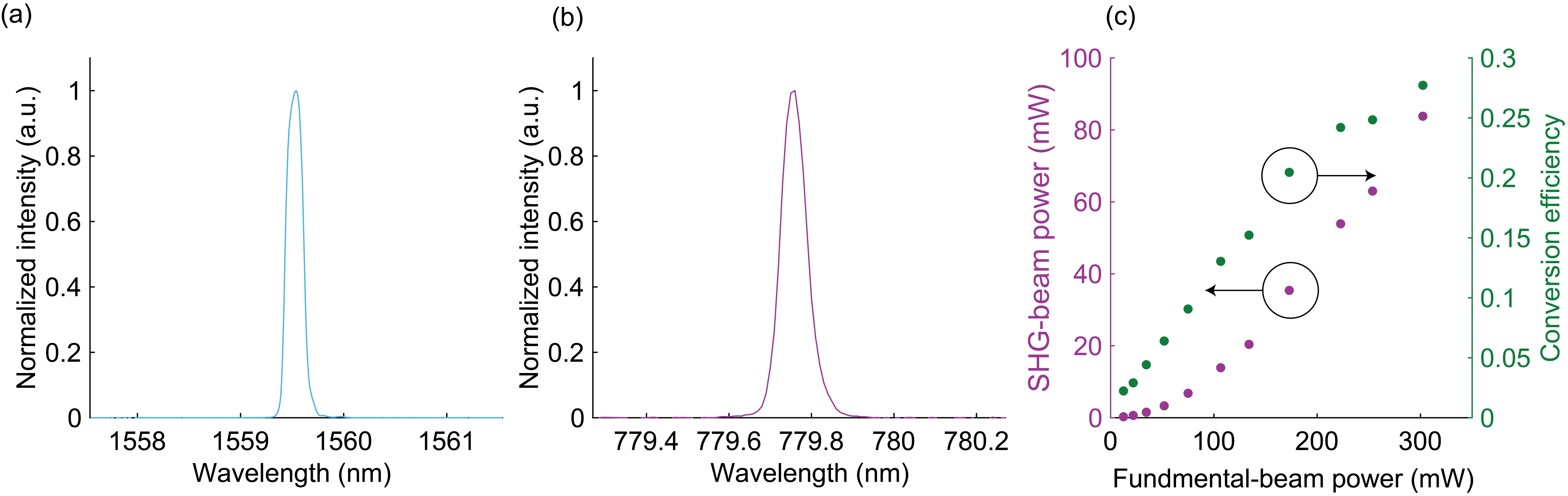}
\caption{Fundamental and SHG beam characterization.
 (a) Fundamental and (b) SHG spectra measured by an OSA with a spectral resolution of 0.05 nm.
 (c) SHG power as a function of the fundamental power, together with the corresponding conversion efficiency.}
\label{fig:fun_shg_eval}
\end{figure}

We characterize the spectrum of the fundamental beam after FBG filtering and subsequent amplification with a two-stage amplifier (Figs.~\ref{fig:fun_shg_eval}(a)). The full-width at half maximum (FWHM), measured using an optical spectrum analyzer (OSA) with a resolution of 0.05 nm, is 0.19 nm. Depending on the Er/Yb co-doped fiber amplifier (EYDFA) current, an output power of hundreds of mW is obtained at the input of the 50-mm-long periodically poled lithium niobate (PPLN) crystal used for second-harmonic generation (SHG). The corresponding SHG spectrum is shown in Figs.~\ref{fig:fun_shg_eval}(b). The measured FWHM of 0.07 nm is attributed to the finite instrumental resolution, as the SHG spectral width estimated from the fundamental spectral width is 0.047 nm. 

Figure~\ref{fig:fun_shg_eval}(c) shows the SHG power as a function of the average power of the fundamental beam before the PPLN crystal, along with the corresponding conversion efficiency. The data are obtained by fixing the EYDFA current at 3A and varying the average power of the fundamental beam using an attenuator (ATT). At a fundamental power of 173 mW, an SHG power of 35 mW and a conversion efficiency of 0.2 are achieved. As the average power of the fundamental beam is further increased, the conversion efficiency deviates from linear scaling and begins to exhibit saturation. Since only SHG powers less than around 25 mW are required before the PPLN waveguide used for the optical parametric amplification (OPA) process to generate twin beams, the available SHG power is sufficient for the present implementation. In the measurements presented in the main text and Supplementary Information, the EYDFA current is set to 1.9 A.

\section*{Supplementary Note 4: Characterization of twin beams}
\begin{figure}[htbp]
\centering
\includegraphics[width=0.8\linewidth]{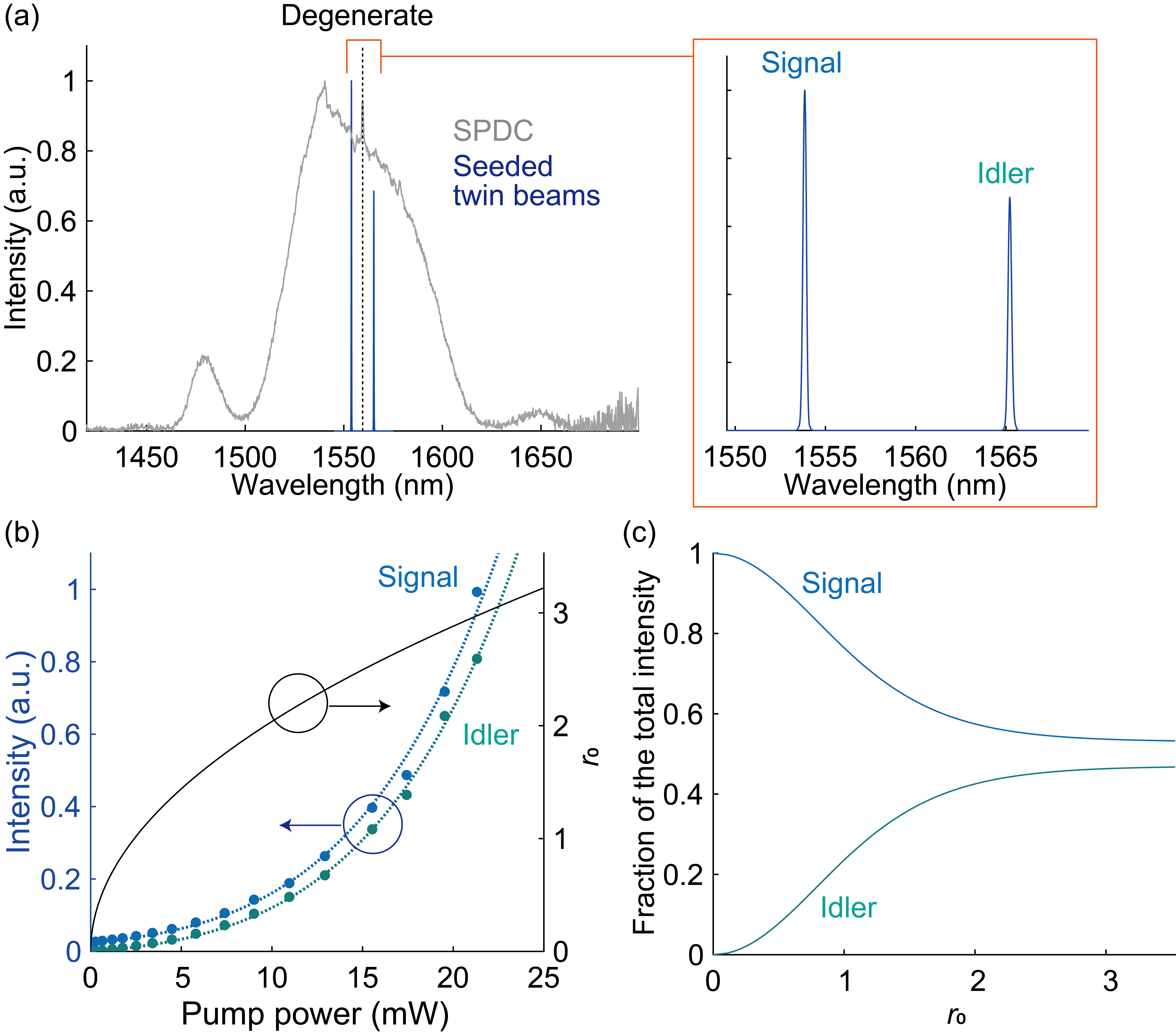}
\caption{Seeded twin-beam characterization. (a) SPDC spectrum measured using an OSA with a spectral resolution of 1 nm at a waveguide temperature of 37.2 $^o$C (gray), together with twin-beam spectra generated via an OPA seeded with a 1553.86-nm coherent beam measured using the OSA with a spectral resolution of 0.05 nm (blue). Each plot is normalized to its respective maximum value. The inset shows an enlarged view of the twin beams.(b) Signal (blue) and idler (blue-green) spectral intensities as a function of the pump power, the corresponding fitted curves (dotted lines), and the estimated $r_0$ vs. pump power (black). (c) Signal (blue) and idler (blue-green) fractions of the total intensity as a function of $r_0$.
 }
\label{fig:twin_eval}
\end{figure}

Next, the twin beams generated via the seeded OPA process are characterized. Figure~\ref{fig:twin_eval}(a) shows the spectrum generated by the spontaneous parametric down-conversion (SPDC) process in the PPLN waveguide at a temperature of 37.2$^o$C, measured using an OSA with a spectral resolution of 1 nm. The pump power measured before the waveguide input is 13 mW, and the resulting spectrum is centered at the degenerate wavelength of 1560 nm with an FWHM of 69 nm. The twin beams are generated within this SPDC bandwidth. The spectrum of twin beams generated with a seed wavelength of 1553.86 nm is also shown in Figs.~\ref{fig:twin_eval}(a), measured using an OSA with a resolution of 0.05 nm. The corresponding signal–idler wavelength combination is used for sub-shot-noise MIP measurements in the main text.

Figure~\ref{fig:twin_eval}(b) shows the intensities of the signal and idler beams as a function of the SHG (OPA pump) power, together with the squeezing parameter extracted from a fit to the idler data. This measurement is the same as in Fig. 2 of the main text, but the seed wavelength differs. The fitting function for the idler is $y = A_i\mathrm{sinh}^2 (B \sqrt{x})$, where $A_i$=0.0085,  and $B$=0.6443 are the fitting coefficients, assuming that the mean photon number of the seed beam is sufficiently larger than 1. The signal results are fitted by $y = A_s\mathrm{sinh}^2 (B \sqrt{x})+C_s$, where $A_s$=0.0096  and $C_s$=0.0248 are the fitting coefficients. Ideally, the signal intensity is expected to follow $y=A_i\cosh^2(B\sqrt{x})=A_i\sinh^2(B\sqrt{x})+A_i$. In practice, however, better agreement with the experimental data is obtained by treating the coefficient of the $\sinh^2(B\sqrt{x})$ term and the constant offset as independent fitting parameters. This discrepancy may arise from departures from the ideal model, such as imperfect mode overlap, whereby only a fraction of the seed effectively participates in the OPA process. To determine the squeezing parameter $r_0$, we use the value of $B$ obtained by fitting the idler intensity and calculate $r_0$ as $r_0=B\sqrt{x}$. Figure~\ref{fig:twin_eval}(c) shows the signal and idler fractions of the total intensity, derived from the fitting results in (b). These plots are used to determine the signal (MIP probe) power, as well as the corresponding mean photon number per pulse, from the measured total twin-beam power.

\section*{Supplementary Note 5: Single-port and balanced detection}
\begin{figure}[htbp]
\centering
\includegraphics[width=0.5\linewidth]{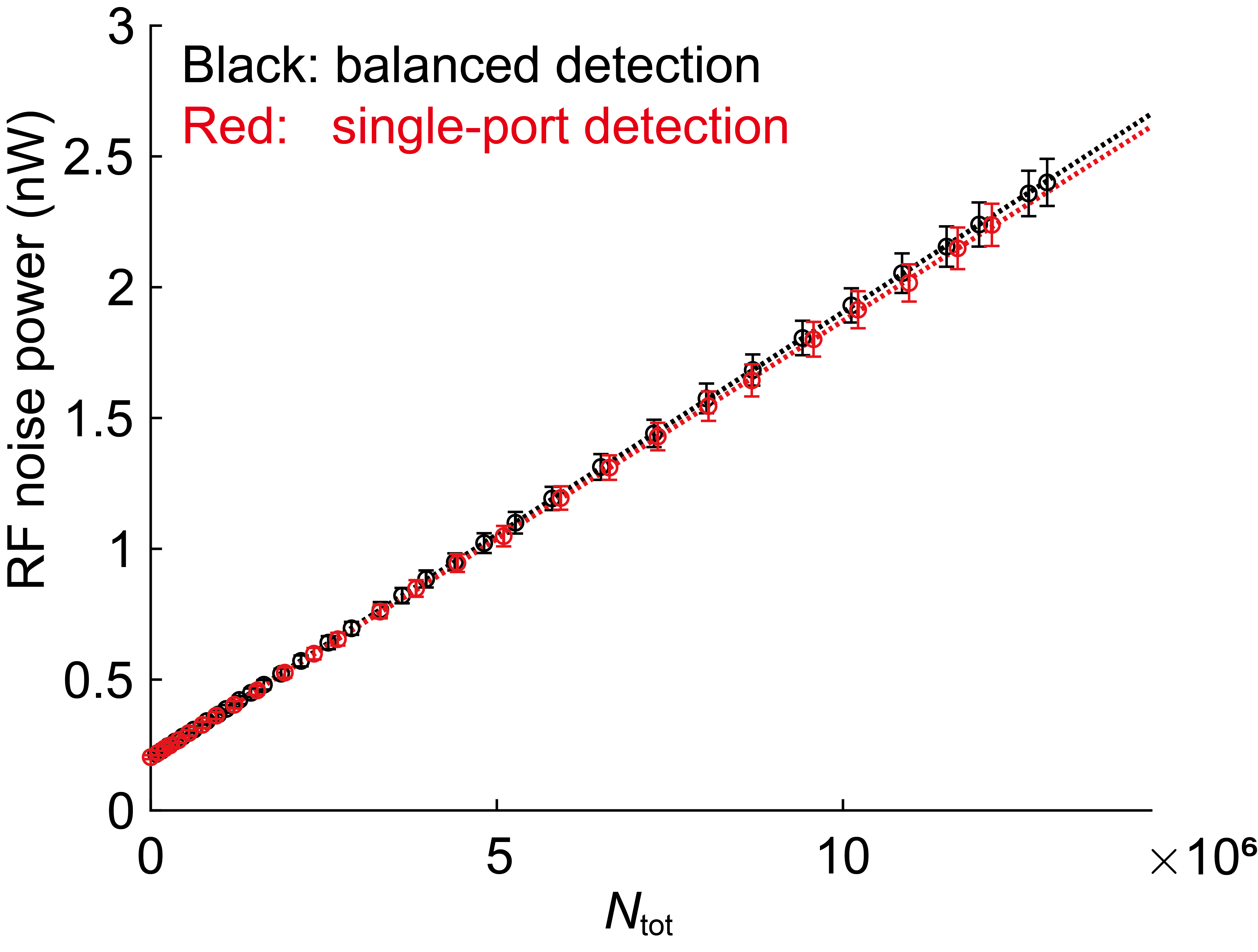}
\caption{RF noise power as a function of $N_{\mathrm{tot}}$ measured with single-port (red) and balanced (black) detection. The corresponding dotted lines represent the linear fitting curves. The y-axis error bars represent the standard deviation of the RF noise power over time, as measured in zero-span mode.}
\label{fig:single_balanced}
\end{figure}

In the evaluation of the NRF in Fig. 3 of the main text and Figs. ~\ref{fig:NRF_twin} and ~\ref{fig:NRF_loss}, we compare the noise performance of a coherent beam detected in a single-port configuration with that of twin beams measured via balanced detection, both normalized to the total mean photon numbers per pulse $N_{\mathrm{tot}}$. For shot-noise-limited measurements, comparing the variance normalized by the total detected photon number yields identical results for single-port detection and balanced detection of a coherent beam. Figure~\ref{fig:single_balanced} shows the measured noise at 5 MHz for a coherent beam (seed beam for OPA), obtained using (i) single-port detection with a band-pass filter (BPF) and (ii) balanced detection with a pellicle beamsplitter(BS). The measurements are performed in zero-span mode with a resolution bandwidth of 1 MHz and a video bandwidth of 1 kHz. In both cases, the RF power scales linearly with $N_{\mathrm{tot}}$, consistent with shot-noise-limited behavior, and the two measurements agree within the experimental uncertainty. These results indicate that the shot-noise curve obtained from single-port detection of a coherent beam can be used to derive the NRF for twin-beam measurements.

\section*{Supplementary Note 6: NRF with several seed photon numbers}
\begin{figure}[htbp]
\centering
\includegraphics[width=1\linewidth]{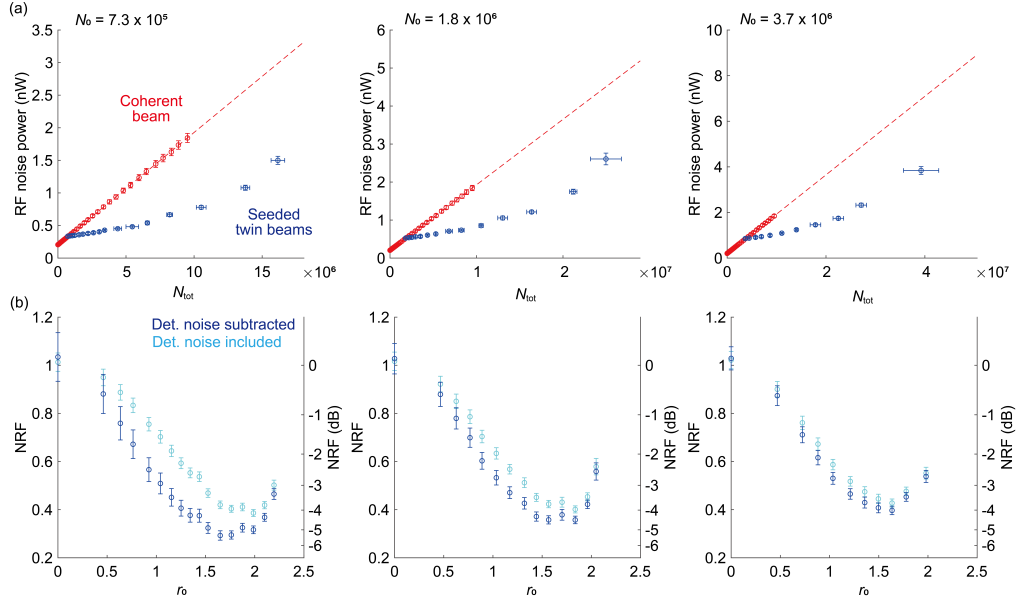}
\caption{NRF measured at several $N_{0}$. (a) RF noise power of the twin beams (blue) with $N_{0}=$ 7.3×10$^5$(left), 1.8 x 10$^6$ (center), and 3.7 x 10$^6$ (right), and their comparison with the coherent beam cases (red). The RF noise power is measured using an RFSA operated in zero-span mode with a center frequency of 5 MHz, a resolution bandwidth of 1 MHz, and a video bandwidth of 1 kHz. The dashed red line indicates a linear fit to the shot-noise level of the coherent beam. The x- and y-axis error bars represent the standard deviations of $N_{\mathrm{tot}}$ and the measured RF noise power, respectively. (b) Corresponding NRFs as a function of $r_0$ with (blue) and without (sky blue) the detector noise subtraction. The y-axis error bars represent the standard deviations of the NRFs obtained by propagating the standard deviations of the RF noise powers shown in (a).}
\label{fig:NRF_twin}
\end{figure}

Figure~\ref{fig:NRF_twin} shows the results of the same measurements as those presented in Fig. 3 of the main text, performed at higher seed mean photon numbers per pulse $N_{0}=$ 7.3×10$^5$, 1.8×10$^6$, and 3.7×10$^6$. In the case of Fig. 3, where $N_{0}=$ 3.6×10$^5$, an NRF better than -5 dB is obtained after subtracting the detector noise. In contrast, at higher seed photon numbers, the minimum achievable NRF degrades as the photon number increases from 1.8×10$^6$ to 3.7×10$^6$ photons per pulse. For the case of $N_{0}=$ 3.7×10$^6$, the best NRF after subtracting the detector noise is -4 dB.

\section*{Supplementary Note 7: NRF under optical loss}
\begin{figure}[htbp]
\centering
\includegraphics[width=1\linewidth]{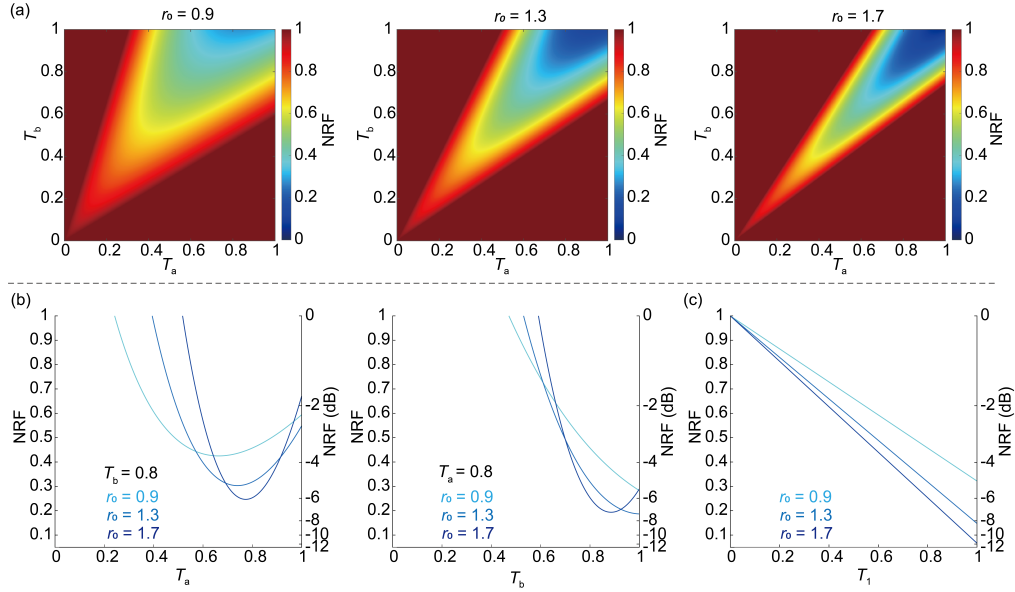}
\caption{Theoretically calculated transmittance dependence of the NRF for seeded twin beams.(a) Heat maps showing the $T_a$ and $T_b$ dependence of the NRF calculated for $r_0=$0.9, 1.3, and 1.7. (b) Left: NRF as a function of $T_a$ for $T_b=$0.8. The sky-blue, blue, and dark-blue curves correspond to the cases of $r_0=$0.9, 1.3, and 1.7, respectively. Right: NRF as a function of $T_b$ for $T_a=$0.8.(c) NRF as a function of $T_1(=T_a=T_b)$.}
\label{fig:NRF_loss_theory}
\end{figure}

\begin{figure}[htbp]
\centering
\includegraphics[width=0.5\linewidth]{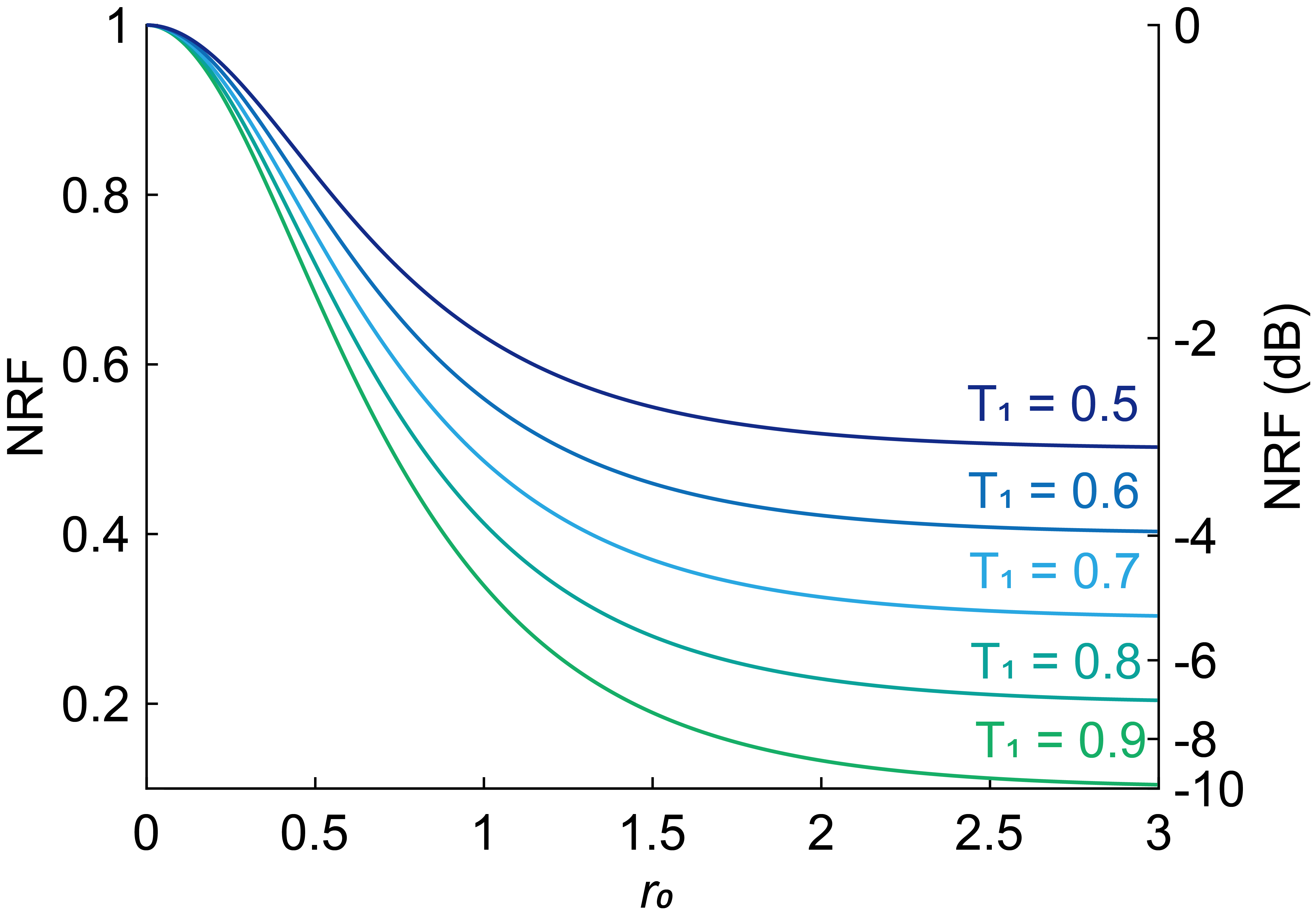}
\caption{Theoretically calculated NRF versus $r_0$ at $T_1=$0.5, 0.6, 0.7, 0.8, and 0.9.}
\label{fig:NRF_loss_theory_gain}
\end{figure}

Based on the derived theoretical expressions, heat maps of the NRF as functions of the transmittances $T_{a}$ and $T_{b}$ are shown in Figs.~\ref{fig:NRF_loss_theory}(a) for $r_0=$0.9,1.3, and 1.7. Increasing $r_0$ reduces the minimum achievable NRF, but it also makes the NRF more susceptible to asymmetric optical loss, such that introducing loss in only one of the twin beams leads to a more pronounced degradation of the NRF. Figure~\ref{fig:NRF_loss_theory}(b) shows the dependence of the NRF on the transmittance of one arm while fixing the other, corresponding to horizontal or vertical cross-sections of the heat maps in (a). When examining the NRF response to variations in signal transmittance, a larger $r_0$ yields a lower minimum NRF but also makes it more susceptible to loss, resulting in more rapid NRF degradation. A similarly steep degradation is observed when varying the idler transmittance while fixing the signal transmittance. Figure~\ref{fig:NRF_loss_theory}(c) shows the NRF as a function of equal variations in the signal and idler transmittances for $r_0=$0.9,1.3, and 1.7, corresponding to diagonal cross-sections of the heat maps in (a). In this case, the NRF is linearly dependent on the transmittance and is less sensitive to loss than in the asymmetric-loss case. These results indicate that when optical loss is introduced into one beam, applying a loss to the other beam is essential for suppressing NRF degradation. A similar discussion about the loss effects on the NRF is provided in the literature~\cite{Mueller2020}. In the sub-shot-noise MIP measurements described in the main text, a variable ATT consisting of a half-wave plate (HWP) and a polarizing beamsplitter (PBS) is implemented in the idler arm to adjust the optical loss. Figure~\ref{fig:NRF_loss_theory_gain} shows the NRF versus $r_0$ at $T_1=$0.5, 0.6, 0.7, 0.8, and 0.9. The NRF decreases as $r_0$ increases in the theoretical model, but in the experiment it begins to increase after a certain amount of gain.

\begin{figure}[htbp]
\centering
\includegraphics[width=1\linewidth]{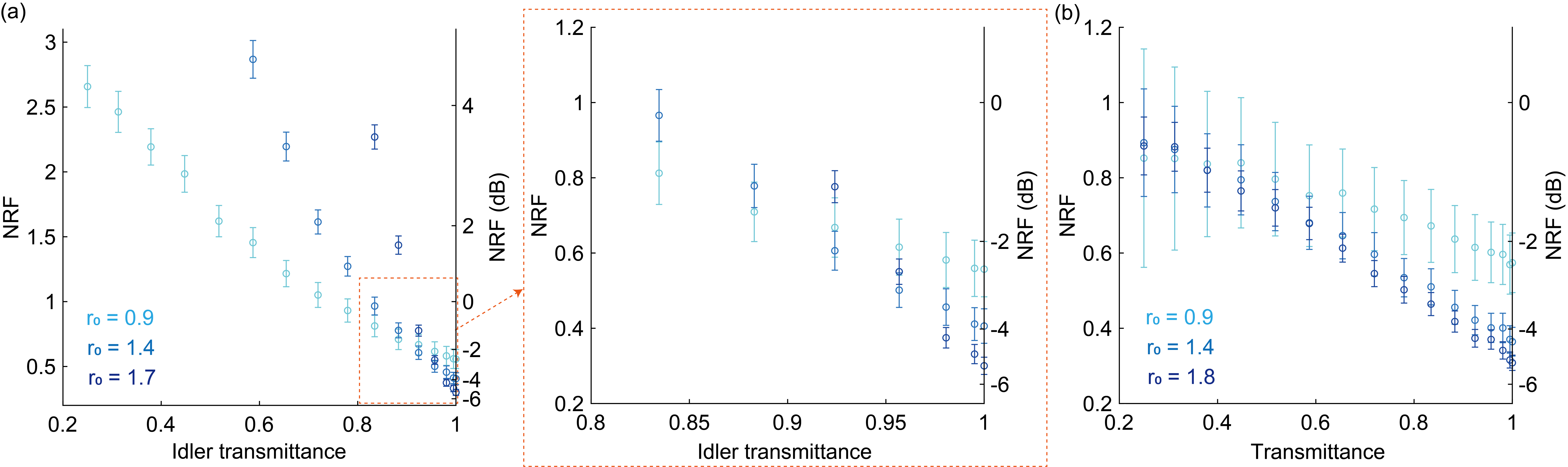}
\caption{Dependence of NRF on optical loss.(a) NRF as a function of the idler transmittance measured at $r_0$=1.7 (dark-blue), 1.4 (blue), and 0.9 (sky-blue). $N_0$ is 3.6 x 10$^5$. The y-axis error bars represent the standard deviations of the NRFs obtained by propagating the standard deviations of the RF noise powers. The inset shows an enlarged view of the panel at an idler transmittance from 0.8 to 1. (b) NRF as a function of the transmittance of the twin beams measured at $r_0$=1.8 (dark-blue), 1.4 (blue), and 0.9 (sky-blue). $N_0$ is 3.6 x 10$^5$. The y-axis error bars indicate the standard deviations of the NRFs calculated from the standard deviations of the RF noise powers.}
\label{fig:NRF_loss}
\end{figure}

Next, we experimentally evaluate the effect of optical loss on the NRF. Figure~\ref{fig:NRF_loss}(a) shows the measured NRF as a function of the idler transmittance for $r_0=$0.9,1.4, and 1.7 with detector noise subtraction. Under the same measurement conditions as in Fig. 3 of the main text, a variable ATT consisting of a HWP and a PBS is inserted in the idler path to vary the idler transmittance. Here, the idler transmittance denotes the relative transmittance referenced to the maximum transmittance of the variable ATT, and is calculated as a function of the HWP's rotation angle. The NRF is then derived by accounting for the change in the total detected optical power due to variations in idler transmittance, and by referencing the shot-noise level. For $r_0=1.7$, an NRF of -5 dB is obtained at the maximum idler transmittance, but the system is vulnerable to loss. For example, when the idler transmittance falls below approximately 0.8, the NRF exceeds 0 dB. Figure~\ref{fig:NRF_loss}(b) shows the effect on the NRF when equal optical loss is applied to both the signal and idler beams by placing the variable ATT at a point where the signal and idler co-propagate, for $r_0=$0.9,1.4, and 1.8. Here, the transmittance denotes the relative transmittance referenced to the maximum transmittance of the variable ATT. The NRF is calculated by accounting for the change in total optical power with transmittance and by referencing the shot-noise level. Consistent with the theoretical prediction, the NRF exhibits a near-linear dependence on the transmittance. For $r_0=$1.8, the NRF remains at -3 dB even at a transmittance of 0.78.

\section*{Supplementary Note 8: MIP with MHz modulation}
We describe how a MIR beam induces a phase shift in the probe beam.

\subsection*{Heat-conduction equation}
The local temperature evolution induced by the photothermal effect is described by the heat conduction equation~\cite{Ishigane2023}:
\begin{equation}
\frac{\partial \Delta T(x,y,z,t)}{\partial t}
=
\alpha_{\mathrm{th}}\nabla^2 \Delta T(x,y,z,t)
+
\frac{I(x,y,z,t)\mu_a(x,y,z)}{\rho C_p},
\label{eq:heat}
\end{equation}
where $x$, $y$, $z$ denote spatial coordinates and $\alpha_{\mathrm{th}}$, $I$, $\mu_a$, $\rho$, $C_p$, $t$, and $\Delta T$ denote the thermal diffusivity, pump intensity, absorption coefficient, density, specific heat capacity, time, and local temperature change, respectively.

\subsection*{Fast modulation approximation}
The modulated components of the mid-infrared (MIR) intensity and temperature change at the frequency $f_{\mathrm{mod}}$ are represented as 
\begin{align}
I_\mathrm{mod}(x,y,z,t)
&=\mathrm{Re}\left[\tilde{I}(x,y,z)e^{i2\pi f_{\mathrm{mod}}t}\right],\\
\Delta T_\mathrm{mod}(x,y,z,t)
&=
\mathrm{Re}\left[
\tilde{T}(x,y,z)e^{i2\pi f_{\mathrm{mod}}t}
\right],
\end{align}
where $\tilde{I}(x,y,z)$ and $\tilde{T}(x,y,z)$ are the corresponding complex amplitudes. Substituting these expressions into Equation~\eqref{eq:heat} and collecting the components oscillating at $f_{\mathrm{mod}}$ yields 
\begin{equation}
\left[ i2\pi f_{\mathrm{mod}}-\alpha_{\mathrm{th}}\nabla^2 \right]\tilde{T}(x,y,z)
=
\frac{\mu_a(x,y,z)}{\rho C_p}\tilde{I}(x,y,z).
\label{eq:heat2}
\end{equation}
The thermal diffusion length~\cite{Xia2022}, determined by $L_{\mathrm{th}} = 2\sqrt{\frac{\alpha_{\mathrm{th}}}{f_{\mathrm{mod}}}}$, is $ 0.54\ \mu\mathrm{m}$ with a water thermal diffusivity of $\alpha_{\mathrm{th}}=1.44\times10^{-7}\ \mathrm{m^2\,s^{-1}}$ and $f_{\mathrm{mod}}=$ 2 MHz. Since $L_{\mathrm{th}}$ is much smaller than the MIR beam diameter in water $2w_{\mathrm{MIR}}\approx32\ \mu\mathrm{m}$, the thermal diffusion term can be neglected. Under these approximations, Equation~\eqref{eq:heat2} reduces to
\begin{align}
\tilde{T}(x,y,z)\approx
-i\frac{\mu_a(x,y,z)}{2\pi f_{\mathrm{mod}}\rho C_p}\tilde{I}(x,y,z). 
\end{align}

\subsection*{MIP phase}
The complex amplitude of the MIP-induced phase modulation is given by
\begin{equation}
\tilde{\phi}(x,y)\approx
\frac{2\pi}{\lambda_{pr}}\frac{dn}{dT}\frac{1}{\cos\theta_{p}}\int \tilde{T}(x+z\tan\theta_{p},y,z)\,dz,
\end{equation}
where $\dfrac{dn}{dT}$, $\lambda_{pr}$, and $\theta_{p}$ denote the thermo-optic coefficient (TOC) of the sample, the probe wavelength, and the refracted angle of the probe beam inside the sample, respectively. Assuming the absorption coefficient is spatially uniform for simplicity, the phase modulation can be expressed as
\begin{equation}
\tilde{\phi}(x,y)\approx
-i\frac{\mu_a}{f_{\mathrm{mod}}\rho C_p \lambda_{pr}}\frac{dn}{dT}\frac{1}{\cos\theta_{p}}\tilde{I}_{z,\theta_{p}}(x,y),
\end{equation}
where $\tilde{I}_{z,\theta_{p}}(x,y)=\int \tilde{I}(x+z\tan\theta_{p},y,z)\,dz.$ The magnitude of MIP phase modulation is proportional to the absorption coefficient, the magnitude of the TOC, and the inverse of the modulation frequency. The photothermal effect induces a localized change in the refractive index of the sample, thereby modifying the wavefront of the probe beam~\cite{Domínguez-Juárez2023}.

\section*{Supplementary Note 9: Fourier image of the probe beam}
\begin{figure}[htbp]
\centering
\includegraphics[width=0.8\linewidth]{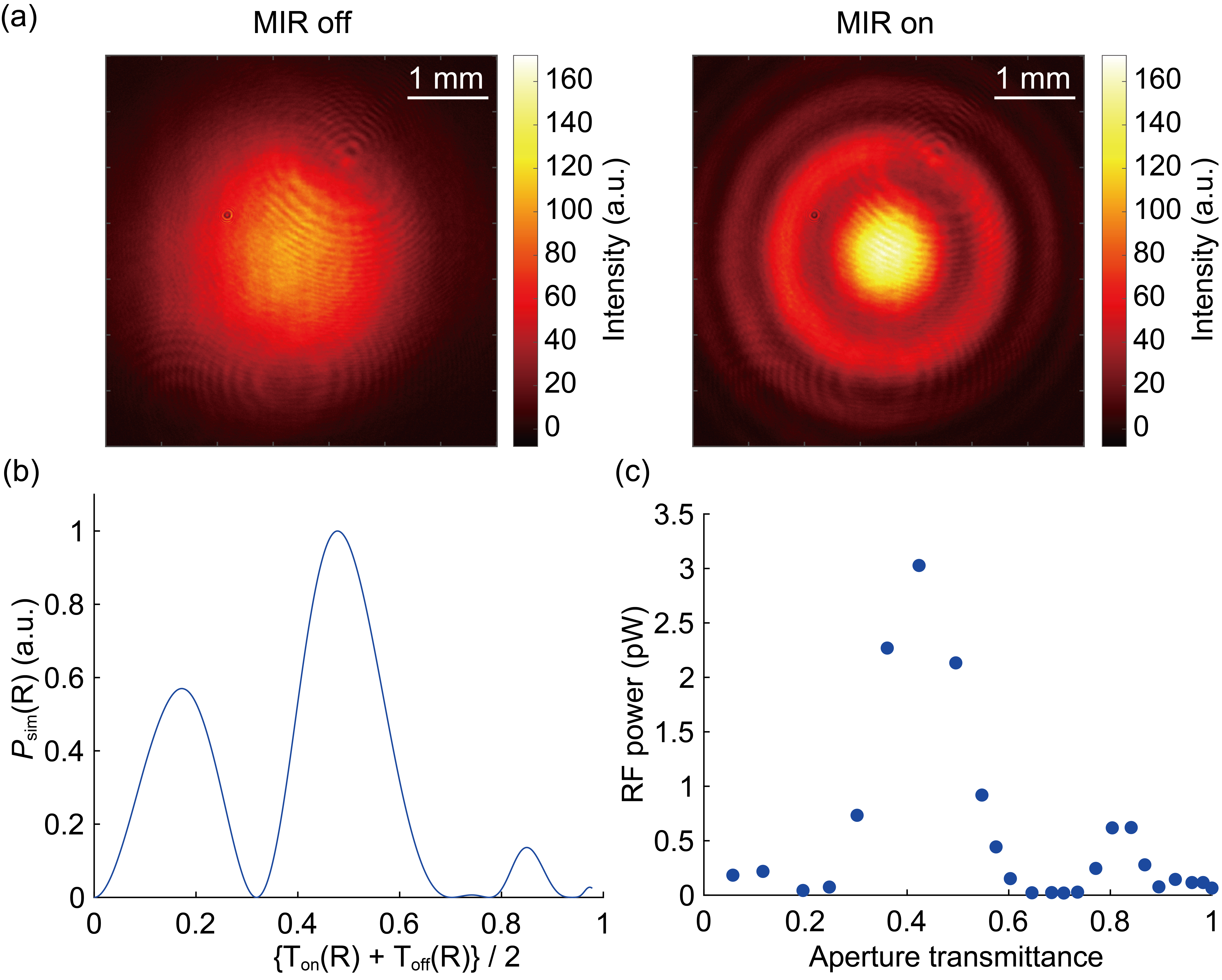}
\caption{Probe beam profile modification induced by the MIP effect.
 (a) Fourier-plane images of the probe beam with and without MIR illumination, measured using an InGaAs camera.
 (b) $P_{\mathrm{sim}}(R)$ as a function of average aperture transmittance under MIR-on and -off conditions: $\{T_{\mathrm{on}}(R)+T_{\mathrm{off}}(R)\}/2$.  
 (c) RF power of the MIP signal at 2 MHz as a function of the aperture transmittance in the Fourier plane. }
\label{fig:beamprofile}
\end{figure}

Figure~\ref{fig:beamprofile}(a) shows the Fourier-plane images with (left) and without (right) the MIR pump measured using an InGaAs camera (STC-SBS322U3V-SWIRU, OMRON SENTECH). Without MIR irradiation, the probe exhibits a Gaussian-like profile, whereas a ring-like pattern is observed under MIR irradiation. 

The MIP signal detected at $f_{\mathrm{mod}}$ is determined by the Fourier-plane intensity modulation. We denote the corresponding complex amplitude by $\widetilde I_{pr}(k_x,k_y)$, where $(k_x,k_y)$ are the transverse spatial-frequency coordinates. The optical-power modulation transmitted through a circular aperture $\mathcal A_{R}$ with a radius of $R$ is
\begin{align}
\widetilde P_{R}
=\iint_{\mathcal A_{R}}
\widetilde I_{pr}(k_x,k_y)dk_xdk_y,
\end{align}
and thus, the RF power measured at $f_{\mathrm{mod}}$ is
\begin{align}
P_{\mathrm{RF}}(R)
\propto\left|\widetilde P_R\right|^2
=\left|\iint_{\mathcal A_R}\widetilde I_{pr}(k_x,k_y)dk_xdk_y\right|^2.
\end{align}
The aperture converts the spatial redistribution of the modulated probe intensity into a modulation of the total detected optical power.

Since the camera does not directly resolve the 2 MHz intensity-modulation component, we use the difference between the time-averaged MIR-on and -off images as an empirical proxy for its spatial distribution. Figure~\ref{fig:beamprofile}(b) shows the resulting aperture-dependent quantity,
\begin{align}
P_{\mathrm{sim}}(R)\propto\left|T_{\mathrm{on}}(R)-T_{\mathrm{off}}(R)\right|^2,
\end{align}
where $T_{\mathrm{on}}(R)$ and $T_{\mathrm{off}}(R)$ are the fractions of the total image intensity transmitted through a circular aperture under the MIR-on and -off conditions, respectively. Note that the procedure is not a direct calculation of the 2 MHz RF power, but it provides an empirical model of its aperture dependence under the assumption that the MIR-on/off image difference reproduces the principal spatial features of the modulation component. The ring-like redistribution of the probe intensity causes the integrated transmitted power to vary non-monotonically with aperture size, resulting in multiple local maxima. Figure~\ref{fig:beamprofile}(c) shows the experimentally measured RF peak power of the 2 MHz MIP signal as a function of aperture transmittance. The experimentally observed multiple maxima are qualitatively consistent with the behavior predicted from the MIR-on/off image difference. In the sub-shot-noise MIP experiments presented in the main text, the aperture is set near the local maximum at a transmittance of 0.8 to preserve the noise reduction advantage of the twin beams.

\section*{Supplementary Note 10: MIP measurements with twin beams}
\begin{figure}[htbp]
\centering
\includegraphics[width=0.5\linewidth]{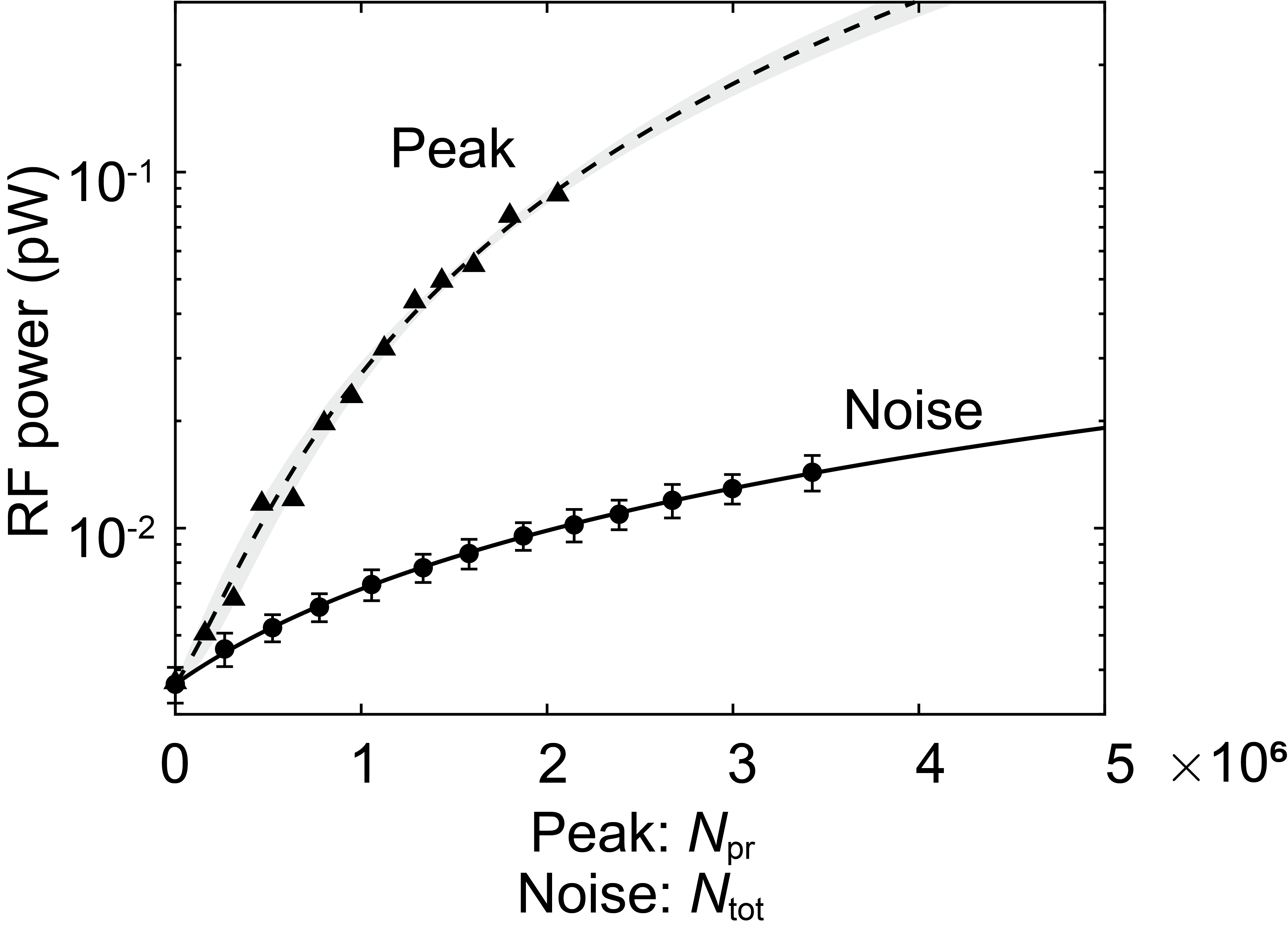}
\caption{Peak (triangle) and noise (circle) powers as a function of $N_{\mathrm{pr}}$ and $N_{\mathrm{tot}}$, respectively, in balanced-detection MIP measurements using coherent beams. The dashed and solid curves represent the corresponding fitting results for the peak and noise, respectively, with their shaded areas representing fitting errors. The y-axis error bars for the noise represent the standard deviation of the RF noise power measured over the frequency range from 1.9990 to 1.9998 MHz.}
\label{fig:twinbeams_shot}
\end{figure}

The fitting curves in Fig. 6 of the main text are obtained from the corresponding measurements with balance detection of a coherent beam (Figs.~\ref{fig:twinbeams_shot}). The dashed and solid lines represent a quadratic fit for the peak power and a linear fit for the noise power, respectively. The shaded gray areas represent the corresponding fitting errors for the peak and noise powers. 

\begin{figure}[htbp]
\centering
\includegraphics[width=0.8\linewidth]{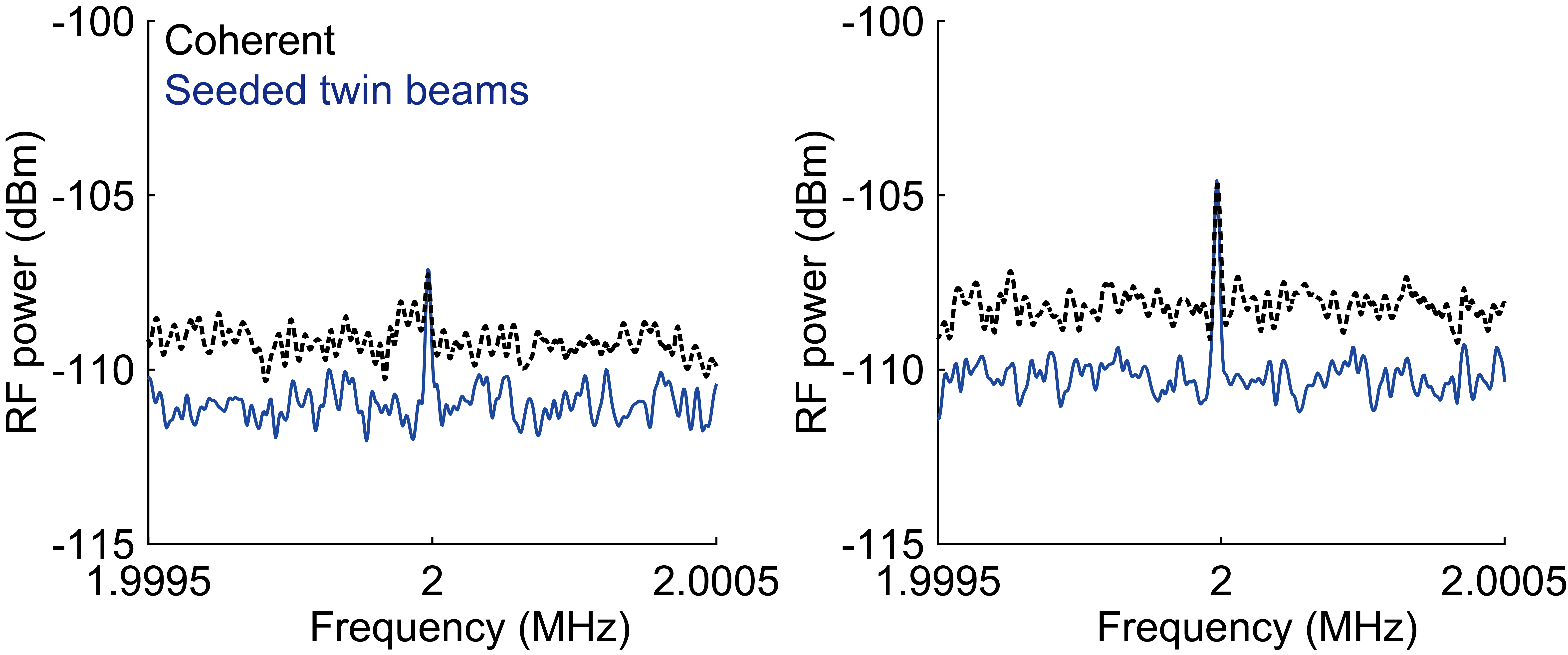}
\caption{MIP measurements with twin beams at an aperture transmittance of 0.93. Left: RF spectra measured using the twin beams with $r_0=$ 1.8 (blue) ($N_{\mathrm{tot}}=$ 2.8x10$^6$, $N_{\mathrm{pr}}=$ 1.8x10$^6$) and a coherent beam (dotted black)($N_{\mathrm{tot}}=$ 2.7x10$^6$, $N_{\mathrm{pr}}=$ 1.4x10$^6$). Right: RF spectra measured using the twin beams with $r_0=$ 2.0 (blue) ($N_{\mathrm{tot}}=$ 4.0x10$^6$, $N_{\mathrm{pr}}=$ 2.6x10$^6$) and a coherent beam (dotted black)($N_{\mathrm{tot}}=$ 3.8x10$^6$, $N_{\mathrm{pr}}=$ 2.0x10$^6$).}
\label{fig:RF_coh_twin}
\end{figure}

We also present the measurements shown in Fig. 5 in the main text at a higher transmittance of 0.93 (with lower MIP peak powers) in Figs.~\ref{fig:RF_coh_twin}. In the left panel of Figs.~\ref{fig:RF_coh_twin}, the signal obtained with a coherent beam is close to the noise level. In contrast, for a comparable $N_{\mathrm{tot}}$, the signal peak measured with twin beams is clearly resolved owing to the reduced noise floor.

\begin{figure}[htbp]
\centering
\includegraphics[width=0.8\linewidth]{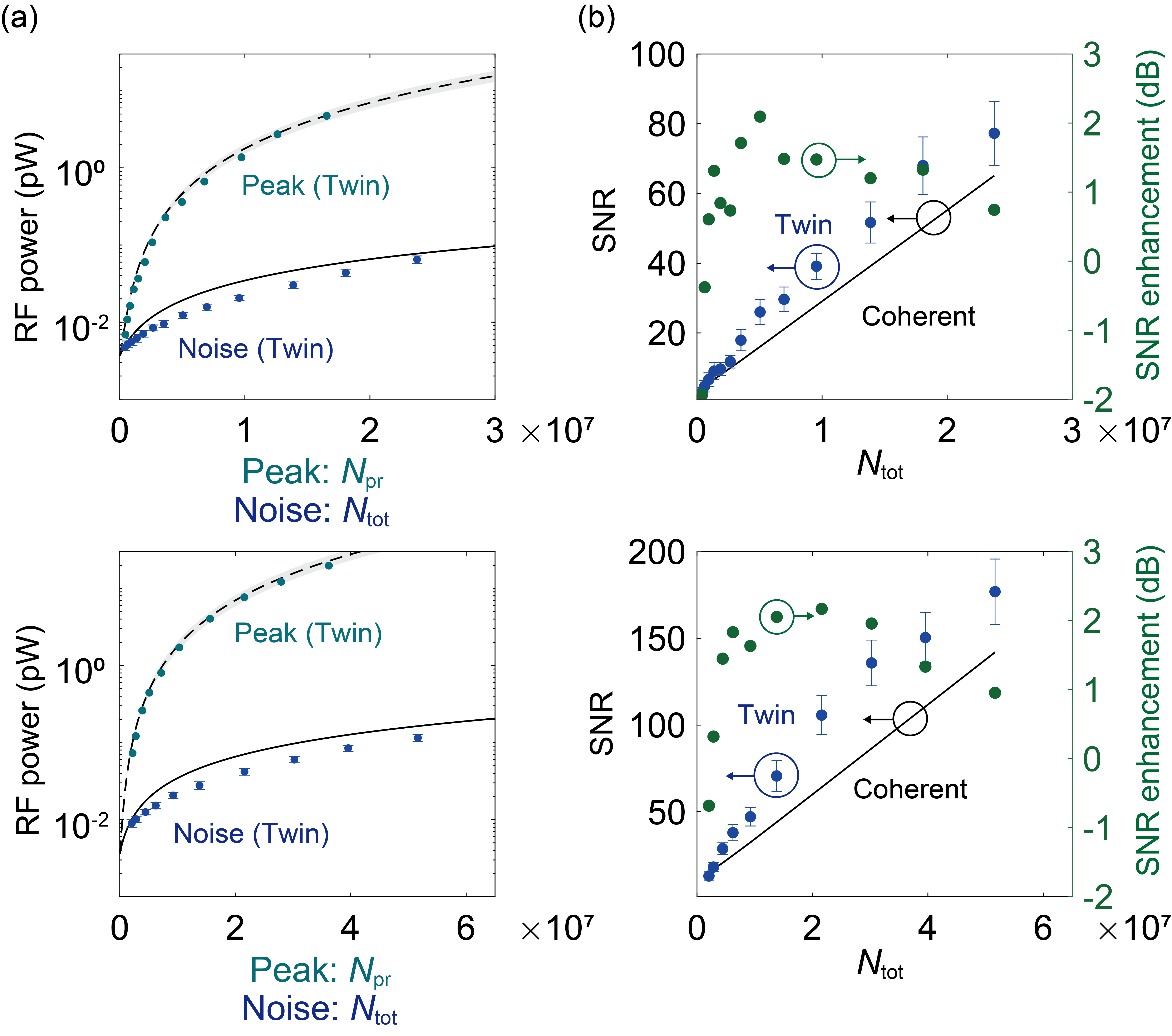}
\caption{Sub-shot-noise MIP measurements using seeded twin beams with $N_0=$ 3.7x10$^5$ (upper) and 1.8x10$^6$ (lower). (a) RF peak (blue-green) and noise floor (blue) for the seeded twin beams as functions of $N_{\mathrm{pr}}$ and $N_{\mathrm{tot}}$, respectively. The y-axis error bars for the noise represent the standard deviation of the RF noise power measured over the frequency range from 1.9990 to 1.9998 MHz. The dashed and solid black lines represent the fitting curves for the peak and noise floor measured using a coherent beam with balanced detection, respectively. The shaded gray areas represent the fitting errors for the dashed and solid black lines. (b) SNR of the twin-beam measurements (blue) as a function of $N_{\mathrm{tot}}$, compared with the coherent-beam case (black). The y-axis error bars indicate the standard deviation of the SNR calculated from the standard deviations of the noise (shown in (a)) using the error propagation law. The green plots show the SNR enhancement achieved using twin beams, calculated as the ratio of the SNR obtained with twin beams to that obtained with coherent beams.}
\label{fig:subshotnoise_si}
\end{figure}

Figure~\ref{fig:subshotnoise_si} presents the results of sub-shot-noise MIP measurements performed under the same conditions as those used for Fig. 6 in the main text, except that the seed mean photon number was set to $N_0=$3.8x10$^5$ and 1.8x10$^6$.

\clearpage

\bibliography{sample}